\documentclass[pra,aps,preprint]{revtex4}
\usepackage{graphicx}

\begin{document}
\title{Quantum Decoherence from Adiabatic Entanglement}
\author{\textsc{C.P.Sun}$^{1,2}$,\textsc{{D.L.Zhou}$^{1}$ , S.X.Yu$^{1}$ and X.F.Liu$^{3}$ }}

\address{$^{(1)} $ Institute of Theoretical Physics, Chinese Academy of Sciences,
Beijing 100080, China\\ $^{(2)} $Department of Physics,The Chinese
University of Hong Kong, Hong Kong, China\\ $^{(3)} $ Department
of Mathematics, Peking University,Beijing, 100871,China}

\begin{abstract}
In order to understand quantum decoherence of a quantum system due
to its interaction with a large system behaving classically, we
introduce the concept of adiabatic quantum entanglement based on
the Born-Oppenhemeir approximation. In the adiabatic limit, it is
shown that the wave function of the total system formed by the
quantum system plus the large system can be factorized as an
entangled state with correlation between adiabatic quantum states
and quasi-classical motion configurations of the large system. In
association with a novel viewpoint about quantum measurement ,
which has been directly verified by most recent experiments
[e.g, S. Durr et.al, Nature 33, 359 (1998)], it is shown that the
adiabatic entanglement is indeed responsible for the quantum
decoherence and thus it can be regarded as a ``clean" quantum
measurement when the large system behaves as a classical object.
The large system being taken respectively to be a macroscopically
distinguishable spatial variable, a high spin system and a
harmonic oscillator with a coherent initial state, three
illustrations are present with their explicit solutions in this
paper.
\end{abstract}

\pacs{03.65.Bz, 03.67.-a}

\maketitle

\section{Introduction}

In quantum measurement process, wave packet collapse (WPC, also called von
Neumann's projection or wave function reduction ) physically resembles the
disappearance of interference pattern for Young's two-slit experiment in the
presence of a ``which-way'' detector. Associated with the wave-particle
duality, this phenomenon of losing quantum coherence is referred to as the
so-called quantum decoherence. In fact, before a measurement to observe
``which-way'' the particle actually takes, the quantum particle seems to
move from a point to another along several different ways simultaneously.
This just reflects the wave feature of a quantum particle. The detection of
``which-way'' means a probe for the particle feature, which leads to the
disappearance of wave feature or quantum decoherence[2].

Further explanation for decoherence phenomena was made in the view point of
complementarity by Niels Bohr based on Heisenberg's position-momentum
uncertainty [3]: A particular interaction between a classical instrument
(detector) and the measured quantum system can be regarded as a quantum
measurement; but once enough data about the states of the quantum system is
``read out'' from the motion configuration of the detector, the interaction
unavoidably destroys the interference pattern. According to Heisenberg's
uncertainty principle, to locate the position of a particle to the
uncertainty of order $\Delta x$ along the direction orthogonal to its moving
direction, the ``which-way'' measurement must kick the momentum of the
particle to an uncertainty of order $1/\Delta x$ and thereby washes out the
spatial interference pattern. (In this paper the Plank constant is taken to
be unity). Bohr's argument sounds correct, but a recent experiment [4] on
Brrag's reflection of cold atoms shows that Schrodinger's concept of
entangled state, rather than the unavoidable measurement distribution, is
crucial for the wave-particle duality in this ``which-way'' experiment.
Another ``which-way'' experiment [5] , which uses the electron Aharonov-Bohm
interference with a quantum point contact, also manifests the importance of
quantum entanglement. Actually, similar \textit{gedenken }experiments using
photon and neutron have been considered before [6-7]

A quantum entangled state [8-10] such as $$ |\Psi \rangle
=\sum_nC_n|S_n\rangle \otimes |D_n\rangle \quad (\neq |S\rangle
\otimes |D\rangle \eqno{(1.1)}$$ for any $|S\rangle $ and
$|D\rangle$ is a coherent superposition of states of a quantum
system of many particles or of a single particle with many degrees
of freedom. It involves a correlation between the states
$|S_n\rangle $ of the quantum system and the states $|D_n\rangle $
of the detector. Once the detector is found in a state
$|D_n\rangle ,$ the total system must collapse into a certain component $%
|S_n\rangle \otimes |D_n\rangle $. Then one can infer the state $|S_n\rangle
$ of the quantum system. The interference pattern can be obtained from the
total wave function $|\Psi \rangle $ by ``summing over" all possible states
of the detector. Assuming the states of the detector are normalized, we have
$$
\sum_m|\langle D_m|\otimes \langle x|\Psi \rangle
|^2=\sum_n|C_n|^2|S_n(x)|^2+\sum_{n\neq m}C_m^*C_nS_m^{*}(x)S_n(x)\langle D_m|D_n\rangle %
\eqno{(1.2)}
$$
where $S_n(x)=\langle x|S_n\rangle $ is the state of the quantum system in
position representation. The second term on the r.h.s of the above equation
is responsible for the interference pattern. It is easy to see the
interference fringes completely vanish when the states of the detector are
orthogonal to one another [10], i.e., when $\langle D_m|D_n\rangle =\delta
_{m.n}$. In this situation, an ideal quantum measurement results from the
ideal entanglement with the orthogonal correlated components $|D_n\rangle $
, in which one can distinguish the state of detector very well.
Mathematically, by using the reduced density matrix
$$
\rho =\mbox{Tr}_D(|\Psi \rangle \langle \Psi |)=\sum_n|C_n|^2|S_n\rangle \langle
S_n|+\sum_{m\neq n}C_m^*C_n|S_n\rangle \langle S_m|\langle D_m|D_n\rangle %
\eqno{(1.3)}
$$
which is obtained by tracing out the variables of detector, the
above-mentioned decoherence phenomenon can be equivalently expressed as a
projection or reduction of the reduced density matrix from a pure state $%
\rho =\sum_{m,n}|S_n\rangle \langle S_m|$ to a mixed state $\hat{\rho}%
=\sum_n|S_n\rangle \langle S_n|.$

It is noticed that, so long as the `` which-way'' information already stored
in the detector \textit{could be read out}, the interference pattern has
been destroyed \textit{without any data read out in practice }[4,5]. In this
sense the environment surrounding the quantum system behaves as a detector
to realize a ``measurement-like '' process. This is because the environment
\textit{never needs to read out} the data. Thus, the above argument is also
applicable to the analysis of decoherence problem of an interfering quantum
system coupling to the environment [8-10].In this kind of problems, the
environment is imagined as an objective detector detecting the states of the
quantum system and thereby the detector states $|D_n\rangle $ are thought to
be the macroscopic quantum states of the environment. Provide the
environment couples with the quantum system and produce an ideal
entanglement, the quantum system must lose its coherence. It is worthy to
point out that this simple entanglement conserves the energy of the quantum
system while destroying the quantum coherence. The loss of energy of the
quantum system can be separately discussed in the quantum dissipation theory
well developed in recent years [11-16]

In our previous works on quantum measurement theory[17-23] , we investigate
how an ideal entanglement appears in the macroscopic limit that the number $%
N $ of particles making up the detectors approaches infinity. It was found
that the\textit{\ factorization structure}

$$
F_{m,n}=\langle D_m|D_n\rangle \equiv \prod_{j=1}^N\ \langle
D_m^{[j]}|D_n^{[j]}\rangle \eqno{(1.4)}
$$
concerning the overlapping of detector-states plays a crucial part in
quantum decoherence. Here, $|D_n^{[j]}\rangle $ are the single states of
those blocks constituting the detector,and $F_{m,n}$ is called decoherence
factor. Since each factor $\langle D_m^{[j]}|D_n^{[j]}\rangle $ in $F_{m,n}$
has a norm less than unity, the product of infinite such factors may
approach zero. This investigation was developed based on the Hepp-Coleman
mode and its generalizations[24-27]. In 1998, this theory was applied to the
analysis of the universality [28] of the environment influences on quantum
computing process [29-31].Parallelly, the classical limit that certain
quantum numbers ( such as angular momentum) are huge is also investigated in
our previous works .

However we have not got a totally-satisfactory answer to the question why
the large system entangling with the small system behaves so classically in
such limit situations. In fact,concerning the transition of the detector
from quantum status to classical status , there were only some vague
presentations[17, 19, 21,] . In a general situation the classical feature of
the large system can not simply be characterized by large quantum numbers,
and thus what is responsible for the classical feature remains unclear yet.
Besides, all of our previous discussions about quantum decoherence are based
on interaction of particular forms, namely the non-demolition interaction
[3]. In this paper, using Born-Oppenheimer (B-O) approximation [32], we
universally consider the decoherence problem for a quantum system coupling
to a large system through a general interaction. This basic approach can be
applied to analyzing influences exerted by environment and detector as well.
Our discussion is also involved with a fundamental problem that the
physicist can not avoid completely : how does the time reversal symmetry
implied by the Schrodinger equation on the microscopic scale turn into the
time reversal asymmetry manifested by quantum decoherence or quantum
dissipation on the macroscopic scale?

This paper is organized as follows. We describe in Sec.2 the adiabatic
factorization of slow and fast dynamic variables in terms of the B-O
approximation and show how the interaction of the large object with a
quantum system causes a quantum entanglement dynamically. In
Sec.3,incorporating the semi-classical approach to the quasi-classical
motion of slow variable in a smooth potential, we manifest that, driven by
the adiabatically-effective Hamiltonians, the final states of the large
object initially in an appropriate state are orthogonal to one another,and
their entanglement with the quantum system leads to decoherence. In Sec.4,
Sec.5 and Sec.6, the universal formalism in Secs.2 and 3 is illustrated by
three explicit examples : \textit{a}. A particle with spin $\frac 12$ moves
slowly in an inhomogeneous magnetic field of varying direction; \textit{b}.
A two level quantum system interacting with a very large spin; \textit{c}. A
quantized cavity field is coupled with a simple harmonic oscillator. The
first illustration is similar to the Stern-Gerlach experiment [33]. It
stresses that the classical properties of the large system can be understood
in terms of the macroscopic distinguishability of its quantum states. The
second illustration reflects a simple presentation of quantum- classical
transition when the quantum number is huge [34]. The third illustration has
certain practical significance as it is relevant to the problem of detecting
gravitational wave by intracavity dynamics [35,36]. It demonstrates the
necessity of choosing a quasi-classical initial state of the large system to
realize the quantum coherence of the quantum system. In Sec.7, based on the
adiabatic approach for quantum decoherence, we discuss the spatial
localization of the macroscopic object resulting from the adiabatic
entanglement between its collective coordinate and the dynamic variables of
particles constituting it. This study provides us with a possible solution
to the Schroedinger cat paradox.Concluding remarks are given in the end.It
includes a brief discussion about the development of quantum dynamic theory
of decoherence. In connection with the coupled channel theory, one of whose
concrete realization is the B-O approach, our discussion reveals the
possibility of generalizing our present work to the case with a complicated
and hence more practical interaction than the over-simplified interaction as
shown in the two examples presented in Sec. 4 , 5 and 6.

\section{Quantum Entanglement via Born-Oppenheimer Approach}

In a very wide sense, any interaction between two quantum systems can cause
an entanglement between them. In general, it then realizes a quantum
measurement in a certain meaning. This is because one quantum system in
different states can act on another with different effects correspondingly.
However, this entanglement and its relevant quantum measurement is generally
not very ideal because the usual interaction can not produce a one-one
correspondence between the states of the two systems. Indeed, only a very
particular interaction or its effective reduction can lead to an ideal
entanglement and thereby an ideal quantum measurement. Nevertheless,
fortunately , so long as one of the two systems can be separated \textit{%
adiabatically} and behaves \textit{classically},as we will prove in the
following, any interaction can result in an ideal entanglement in the
evolution of the total system through its adiabatic reduction based on
Born-Oppenheimer (B-O) approximation.

From the view point of BO approximation, we consider a total quantum system
(``molecular'') with two sets of variables, a fast (``electric'') one $q$
and a slow (nuclear) one $x$. Resolving the dynamics of fast variables for a
given motion of the slow subsystem, we obtain certain quantum states labeled
by $n$ for the fast part. To the first order approximation, the left
effective Hamiltonian governing the slow variables involves an external
scalar potential $V_n(x)$ and an magnetic-like vector potential $A_n(x)$
induced by the fast variables [37,38]. The latter is called the induced
gauge potential or Berry's connection. If we assume the motions of the slow
subsystem are ``classical'', we naturally observes that, due to the
back-actions of the fast part , there are different induced forces
$$
F_n=-\nabla _xV_n(x)+\frac d {dt}{x}\times (\nabla _x\times A_n)\eqno{(2.1)}
$$
exerting on the slow part. Their direct physical effects are that the
information of the ``fast'' states labeled by $n$ is recorded in the
different motion configurations of the slow part. An entanglement just stems
from this correlation between the quantum states of the fast subsystem and
the classical motion configurations of the slow subsystem. In spirit of this
physically-intuitive observation, we study the production of such quantum
entanglement from the adiabatic separation of slow and fast variables based
on the B-O approach.

Let us consider the interaction between a quantum system $S$ with fast
dynamic variable $q$ and the large system $E$ with slow variable $x.$ The
former with the Hamiltonian $H_s=H_s(q)$ can be regarded as a subsystem
soaked in an environment or a measured system monitored by a detector, and
the latter with the Hamiltonian $H_E=H_E(x)$ as the environment or the
detector accordingly. In general the interaction Hamiltonian is written as $%
H_I=H_I(x,q)$ . For a fixed value of slow variable $x$ of $E$ , the dynamics
of the quantum system is determined by the eigen-equation
$$
\lbrack H_s(q)+H_I(x,q)]|n[x]\rangle =V_n(x)|n[x]\rangle \eqno{(2.2)}
$$
Both the eigen-values $V_n[x]$ and the eigen-state $|n[x]\rangle $ depend on
the slow variable $x$ as a given parameter.

Usually, the variation of the Hamiltonian $H_s(q)+H_I(x,q)$ with $x$ can
cuase transition from an energy level $V_n(x)$ of the quantum system to
another level $V_m(x)$. But within the spatial domain $R$ to which the slow
variable $x$ belongs, if the variable $x$ changes so slowly that the
adiabatic conditions [39-42]
$$
\left| \frac{\langle n[x]|\partial _x|m[x]\rangle \frac d {dt}{x}}{%
V_m(x)-V_n(x)}\right| =\left| \frac{\langle n[x]|\{\partial
_xH_I(x,q)\}|m[x]\rangle \frac d {dt}{x}}{\{V_m(x)-V_n(x)\}^2}\right| \ll 1%
\eqno{(2.3)}
$$
hold for any two of the different energy levels \{$V_n(x)$\} , this
transition can be physically neglected and then the BO approximation works
as an effective approach. Let $|\Phi _{n,\alpha }\rangle $ be the full
eigen- function of the full Hamiltonian $H=H_E(x)+H_s(q)+H_I(x,q)$ for the
total system formed by the large system plus the quantum system. The B-O
approximation treats it as a partially factorized function
$$
\langle x|\Phi _{n,\alpha }\rangle =\phi _{n,\alpha }(x)|n[x]\rangle %
\eqno{(2.4)}
$$
of the slow and fast variables $x$ and $q.$ Here, the set of slow components
$\{\phi _{n,\alpha }(x)=\langle x|\phi _{n,\alpha }\rangle \}$ and the
corresponding eigen-values $\omega _{n,\alpha }$ are obtained by solving the
effective eigen-equation
$$
H_n(x)\phi _{n,\alpha }(x)=\omega _{n,\alpha }\phi _{n,\alpha }(x)%
\eqno{(2.5)}
$$
The effective Hamiltonian $H_n(x)$ is defined by
$$
H_n(x)=H_{nE}(x)+V_n(x)\eqno{(2.6)}
$$
where $H_{nE}(x)$ is an gauge-covariant modification of $H_E(x)$. It was
obtained by replacing the momentum operator $p=-i\hbar \bigtriangledown _x$
with its gauge-covariant form $p=-i\hbar \bigtriangledown _x-A_n(x)$ . Here,
$A_n(x)=i\langle n[x]|\bigtriangledown _xn[x]\rangle $ is a $U(1)$ gauge
potential induced by the motion of the quantum system. In the classical
limit that the slow part behaves classically, an effective dynamics of
interaction between quantum and classical objects naturally results from the
effective Hamiltonians or its relevant Lagrangian[43].

The completeness relations $\sum_{n,\alpha }|\Phi _{n,\alpha }\rangle
\langle \Phi _{n,\alpha }|=1$for the full eigen-functions $|\Phi _{n,\alpha
}\rangle $ can be expressed in $x-representation$ as
$$
\sum_{n,\alpha }\int dxdx^{\prime }\phi *_{n,\alpha }(x^{\prime })\phi
_{n,\alpha }(x)|x\rangle \langle x^{\prime }|\otimes |n[x]\rangle \langle
n[x]|=1\eqno{(2.7)}
$$
which is equivalent to
$$
\sum_n|x\rangle \langle x|\otimes |n[x]\rangle \langle n[x]|=|x\rangle
\langle x|,\newline
\qquad \sum_\alpha |\phi _{n,\alpha }\rangle \langle \phi _{n,\alpha }|=1%
\eqno{(2.8)}
$$

After obtaining the complete set \{$\phi _{n,\alpha }(x)|n[x]\rangle $\} of
eigenstates of the total system, we can now consider how the entanglement
appears in the adiabatic dynamic evolution. Let the total system be
initially in the state $|\Psi (t=0)\rangle :$
$$
\langle x|\Psi (t=0)\rangle =\{\sum_nc_n|n[x]\rangle \}\phi (x)\eqno{(2.9)}
$$
The first component of the initial state $|\Psi (t=0)\rangle $ is a
superposition of the eigenstates of the quantum system while the second one
a single pure state. Expanding $|\Psi (t=0)\rangle $ in terms of the
complete set \{$\phi _{n,\alpha }(x)|n[x]\rangle $\}, we have the evolution
wave function at time $t$
$$
\langle x|\Psi (t)\rangle =\sum_{n,\alpha }c_n\langle \phi _{n,\alpha }|\phi
\rangle \exp [-i\omega _{n,\alpha }t]|n[x]\rangle \phi _{n,\alpha }(x)%
\eqno{(2.10)}
$$
where we have used the completeness relation eq.(2.7). In terms of the
effective Hamiltonian $H_n(x)$ related to each single adiabatic state $%
|n[x]\rangle ,$ the above wave function is rewritten in a concise form
$$
\langle x|\Psi (t)\rangle =\sum_nc_n|n[x]\rangle \langle x|D_n(t)\rangle %
\eqno{(2.11)}
$$
with
$$
|D_n(t)\rangle =\sum_\alpha \langle \phi _{n,\alpha }|\phi \rangle
e^{-i\omega _{n,\alpha }t}|\phi _{n,\alpha }\rangle =\exp [-iH_nt]|\phi
(x)\rangle \eqno{(2.12)}
$$
The full wave function$|\Psi (t)\rangle $ is obviously an entangled state.
Starting from the same initial state $|\phi \rangle $ at $t=0$, the large
system will be subject to different back-actions defined by $(V_n,A_n)$ from
the different adiabatic states $|n[x]\rangle $ of the quantum system. Then
it evolves to a superposition of different final states $|D_n(t)\rangle $.
This intuitive argument shows us that, there indeed exists an entanglement
between two quantum systems with an quite general interaction,if one of them
moves so slowly that their dynamic variables can be adiabatically factorised
according to the B-O approximation. Roughly speaking, in the B-O approach,
the slow subsystem is usually referred to as heavy particles (such as
nucleons ) while the fast one as light particles (such as the electrons). So
it is reasonable to expect the slow subsystem to behave as a classical
object.

\section{Decoherence: Transition from Quantum to Classical}

In this section we will discuss under what conditions the large system, the
environment or the detector, can behave classically so that the quantum
system entangled with it could completely lose its coherence and approach
the classical limit.

Consider the reduced density matrix of the quantum system
\begin{eqnarray}
\rho _s(t)&=&\mbox{Tr}_D(|\Psi (t)\rangle \langle \Psi
(t)|)=\sum_n|C_n|^2|n[x]\rangle \langle n[x]|+\cr
&&+\sum_{n\neq m}C_mC_n^{*}|m[x]\rangle \langle n[x]|\langle
D_n(t)|D_m(t)\rangle 
\end{eqnarray}

obtained by ''summing over'' the variables of the large system. The
off-diagonal term responsible for interference is proportional to the
overlapping $F_{n,m}=\langle D_n(t)|D_m(t)\rangle $ of the two large system
states.Were there no large system interacting with it, the quantum system
would be completely coherent for $\rho _s(t)=|\varphi (t)\rangle \langle
\varphi (t)|$ is a pure state. Here $|\varphi (t)\rangle =\exp
(-iH_st)|\varphi \rangle $ is a free evolution state of the large system.
Mathematically, the effect of the adiabatic effective interaction is to
multiply the off-diagonal term of the reduced density matrix by the
decoherence factor $F_{n,m}.$ A complete decoherence is defined by $%
F_{n,m}=0 $ while a complete coherence by $F_{n,m}=1(m\neq n).$

Before considering how the decoherence factor $F_{n,m}$ becomes
zero for the large system, we need to review some known arguments
about the meaning of the classical limit of the motion of the
large system. According to a widely accepted viewpoint [44], in
the classical limit, the expectation value of an observable for
certain particular states should recover its classical value
forms. These particular states can give definite classical
trajectories of particle in this limit. Usually we call them
quasi-classical states. A coherent state or its squeezed version
is a typical example of such states. According to Landau and
Lifshitz [44], in general, a quasi-classical state is a particular
superposition $\sum_nc_n\phi _n$ with the non-zero coefficients
$c_n$ only distributing around a large quantum number $
\tilde{n}$ . Then the correspondence principle requires that $%
\tilde{n}\rightarrow \ \infty ,\hbar \rightarrow \ 0$ and the
product $\tilde{n}\hbar $ approaches a finite classical action. In
such a limit, the expectation of an observable will take the
Fourier series of its corresponding classical quantity ; or
strictly speaking, it takes the Fej$\stackrel{^{\prime }}{e}$r's
arithmetic mean of the partial sums of the Fourier series [45]. In
this sense the mean-square deviation of the observable is zero;
and accordingly the mean of the position operator defines a
classical path. Physically, the zero mean-square deviation of the
position operator implies the zero width of each wave packet
$\langle x|D_m(t)\rangle ,$and the overlapping $F_{n,m}=\langle
D_n(t)|D_m(t)\rangle $ of zero width wave packets must vanish.
From such a semi-classical picture, we will clearly see in the
following how the decoherence factor $F_{n,m}$ approaches zero
dynamically as the large system becomes classical.

In the semi-classical approach, for a heavy particle, the initial state $%
|\varphi \rangle $ can be regarded as a very narrow wave packet of width
$a$. Since the heavy particle has a large mass $M\;$ it hardly spreads in the
evolution because without the environment induced quantum dissipation
[15,16] the width of the wave packet at time $t$ is
$$
w(t)=a\sqrt{1+\frac{t^2}{4M^2a^4}}\eqno{(3.2)}
$$
Then we describe the large system as an moving wave packet with the center
along a classical path $x(t)$ on a manifold with local coordinates $x.$ For
a proper initial state $|\varphi \rangle ,$we will see that the wave packet
will split into several narrow peaks with the centers along different paths
determined by different motion equations governed by the effective forces $%
F_n=-\nabla _xV_n(x)+\frac d {dt}{x}\times (\nabla_x\times A_n)$ with
effective potentials $(V_n(x),A_n).$ Usually,the widths of these peaks are
almost of the same order as that of the original wave packet and each peak
is correlated to an adiabatic quantum state $|n[x]\rangle $ for a large
mass. Except for some moments at which the centers of two or more peaks
coincide, these narrow peaks hardly overlaps with one another. In this
sense, the large system starting from a narrow initial state can reach a
superposition of those states orthogonal to one another. Thus we
approximately have $F_{n,m}=0$ in the classical limit for $m\neq n.$

With reference to the useful analysis in ref.[10], we present an explicit
but sketchy calculation to justify the above physically-intuitive
observation about $F_{n,m}=0$ in the classical limit. Assume the large
system to be a heavy particle with very large mass $M$. In the duration $%
\tau $ of the adiabatic interaction with the quantum system, if the
condition $v\tau \approx (\Delta p/M)$ $\tau \ll \Delta x$ holds, the
momentum $p_0$ of the free heavy particle can not be changed notably. Thus
the contributions of the kinetic term and the induced gauge potential can be
ignored in the wave function evolution of the free heavy particle under this
condition. From this consideration we can approximately write down
$$
|D_n(t)\rangle =e^{-iH_nt}|\phi (x)\rangle \propto e^{-iV_n(x)t}|\phi
(x)\rangle \eqno{(3.3)}
$$
The approximation requires that the effective potential $V_n(x)$ is
satisfactorily smooth or the interaction $H_I(x,q)$ is a smooth function of $%
x.$ So we can use
$$
V_n(x)\approx V_n(0)+F_nx;\ F_n\approx \nabla V_n(0)\eqno{(3.4)}
$$
to re-express the decoherence factor
$$
F_{n,m}=\langle \varphi |\exp \Big(-it\delta F(m,n)x\Big)|\varphi \rangle %
\eqno{(3.5)}
$$
Here, $\delta F(m,n)=F_m-F_n$ is the difference of two external forces
exerted by two adiabatic potentials $V_m(x)$ and $V_n(x).$ Then the role of
the back-action of the quantum system on the large system is summing up the
momentum shift by a quantity $\delta F(m,n)t$ with respect to the initial
state $|\varphi \rangle $. Obviously, when the width $\sigma =a^{-1}$ of the
initial wave packet $\langle p|\varphi \rangle $ in the momentum space is
much less than the momentum shift $\delta F(m,n)t$, the large system will
adiabatically evolves into states orthogonal to one another. In fact, if the
initial state is chosen to be a Guassian wave packet $\langle x|\varphi
\rangle =\frac \sigma {\sqrt{\pi }}\exp [-\frac 12\sigma ^2x^2]$ of width $%
\bigtriangleup x=\frac 1\sigma ,$ the decoherence factor is a Guassian
decaying function of time $t$%
$$
F_{n,m}=\exp\left(-\frac{\delta F(m,n)^2}{4\sigma ^2}t^2\right)\eqno{(3.6)}
$$
As the evolution time $t$ approaches infinity or if we have a very narrow
width $\sigma $ , $F_{n,m}\rightarrow \ 0$ and a quantum decoherence results
from the dynamical evolution automatically.

Generally, we consider a system described by $H_n=p^2/2M+V_n(x)$ without the
induced gauge field. Define $x_c=\langle \varphi |x|\varphi \rangle $ and $%
p_c=\langle \varphi |p|\varphi \rangle $ for an initial state $|\varphi
\rangle $. In the classical regime one may expect that the variations $\xi $
$=x-x_c$ and $p_\xi =p-p_c$ are small compared with $x_c$ and $p_c.$%
Accordingly the potential can be expanded as $$ V_n(x)\simeq
V_n(x_c)+V_n^{\prime }(x_c)\xi +\frac12\;V_n^{\prime \prime
}(x_c)\xi ^2.\eqno{(3.7)} $$ So, approximately the Heisenberg
equations of motion become
 $$ \frac d {dt}{x}=\frac p{M},\quad
\frac d {dt}{p}=-V_n^{\prime }(x_c)-V_n^{\prime \prime }(x_c)\xi
\eqno{(3.8)} $$ Sandwiched by the initial state $|\varphi \rangle
$, the above equations turn into the classical equations of motion
$$
\frac d {dt}{x}_c=\frac{p_c}M,\quad \frac d {dt}{p}_c=-V^{\prime }(x_c).%
\eqno{(3.9)} $$ Now we turn to Schr\"{o}dinger's picture. The
evolution of the initial state is governed by $i\hbar \partial
_t|\varphi (t)\rangle =H_n|\varphi (t)\rangle $. Introduce the
following time-dependent translation $$ |\phi(t)\rangle
=\exp\left\{\frac i{\hbar}\Big(\theta(t)+x_cp_\xi
-p_c\xi\Big)\right\}|\varphi (t)\rangle \equiv S(t)|\varphi
(t)\rangle \eqno{(3.10)} $$ where $\theta (t)$ is determined by
$\dot{\theta}_t=p_c^2/2M+V(x_c)$. Then straightforward calculation
gives $$ i\hbar \partial _t|\phi (t)\rangle =\left(\frac{p_\xi
^2}{2M}+\frac12M\omega_t^2\xi^2 \right)|\phi(t)\rangle
\eqno{(3.11)} $$ where $M\omega _t^2=V^{\prime \prime }(x_c)$ .
This exactly describes an oscillator with time-dependent
frequency. The above direct derivation shows that in the
non-inertial frame moving along the classical orbit, every
quasi-classical system looks like a time-dependent oscillator
whose frequency depends on the orbit. This fact is an established
conclusion and illustrated in Fig 1. Actually, it is present in
many textbooks about path integral. But our argument here is based
on a clear physical picture and is applicable to the three
dimensional case after a slight generalization.

Denote by $|0\rangle $ the vacuum state of the harmonic oscillator with
frequency $\omega _0$ which is equivalent to a Gaussian wave packet of width
$\sigma _0^{-1}=\sqrt{2m\omega _0/\hbar }$.Suppose that initially the system
is in the state $S^{\dagger }(0)|0\rangle $, a coherent state whose center
lies at $(x_c(0),p_c(0))$. At time $t$,the center of the wave packet is
obviously at $(x_c(t),p_c(t))$, and it is reasonable to expect that the
width of the wave packet becomes $\sigma _t^{-1}=\sqrt{2m\omega _t/\hbar }$%
,since the frequency of the time-dependent oscillator changes very slowly.
For two different potentials $V_1(x)$ and $V_2(x)$ the macroscopic
distinguishability is ensured when the width sum of the two evolved packets
is less than their orbital difference, that is, when
\[
\sigma _{1t}+\sigma _{2t}\leq |x_{c1}(t)-x_{c2}(t)|
\]
One cannot expect that this condition can always be fulfilled for all time $t
$. The orbital difference is determined by something like $|V_1^{\prime
}-V_2^{\prime }|$ and the width is determined by the second derivative of
the potential. But their relation is not very clear to us at present. What
is clear is,to have $\langle D_1(t)|D_2(t)\rangle =0$ one should require the
points that fail the inequality form a zero measure set. On the other hand,
the adiabatic approximation also imposes some restrictions on the potential.
To clarify the situation further more sophisticated considerations are
needed.

\section{\textbf{From Macroscopic Distinguishability to Decoherence}}

In the context of quantum measurement,a variant of the Stern-Gerlach (SG)
experiment provides an illustration of the above formalism. Quantum
measurement is mutationally an observing process that ``reads out'' the
system states from the ``macroscopically distinguishable'' states of the
detector. As is shown in the above, if the large particle moves slowly
enough, an adiabatic eigen-state of the quantum system will be correlated to
one of the detector states in the B-O approximation. So the adiabatic
correlation
$$
|1[x]\rangle \rightarrow |D_1(t)\rangle ,|2[x]\rangle \rightarrow
|D_2(t)\rangle ,\cdots\cdots |n[x]\rangle \rightarrow |D_n(t)\rangle %
\eqno(4.1)
$$
between the system states $|n[x]\rangle $ and the detector states $%
|D_n\rangle $ defines a quantum measurement. In the classical limit, this
measurement is thought to be ideal for $|D_n\rangle (n=1,2,...)$ are
orthogonal to one another, i.e., $|D_n\rangle $ are shown to be
``classically- or macroscopically distinguishable''. Once the detector is
found in the state $|D_n\rangle $, we can infer that the system is just in
the state $|n\rangle $. In the following we will quantitatively analyze the
dynamical realization of such an adiabatic measurement in a variant of the
SG experiment.

The original Stern-Gerlach (SG) experiment can be considered as a
quantum measurement process detecting the spin states of particles
from their spatial distribution. The WPC or quantum decoherence
can be described in an dynamical evolution governed by the
interaction between the space- and spin- degrees of freedom. In
its variant, a spin-$\frac 12$ particle initially in a certain
superposition state enters an inhomogeneous magnetic field of
amplitude $B(x)$ with \textit{varying} direction
$\mathbf{n}(x)=(\sin \theta \cos kx,\sin \theta \sin kx,\cos
\theta )$ where $\theta $ is fixed. Its configuration is shown in
Fig.2. A simple experiment though it is,it is among the candidates
of experiments proposed to test the Berry's phase or its
corresponding induced gauge field for a neutron in a static
heliacal magnetic field [46, 42 ]. In the usual S-G experiment,
the direction of the magnetic field is along the fixed $x-$axis,
but in our present model the \textit{polarization direction}
\textit{varies} as the position $x$ changes.

The spatial variable is considered to be the slow system while the spin
-variable to be fast as a quantum system. Corresponding to the eigenvalues $%
V_{\pm }(x)=\pm \mu B(x)$, the adiabatic eigenstates of the spin-Hamiltonian
$H_{spin}=\mu B\left( x\right) \mathbf{n(}x)\cdot \mathbf{\sigma }$ are
$$
|\chi _{+}[x]\rangle =\left[
\begin{array}{c}
\cos \frac \theta 2\;e^{-ikx} \\
\sin \frac \theta 2
\end{array}
\right],\quad |\chi _{-}[x]\rangle =\left[
\begin{array}{l}
\sin \frac \theta 2\;e^{-ikx} \\
-\cos \frac \theta 2
\end{array}
\right] 
$$
Here $\mathbf{\sigma =(\sigma }_x,\mathbf{\sigma }_y,\mathbf{\sigma }_z)$ is
the Pauli spin operator and $\mu $ the gyromagnetic ratio. Let the incoming
beam be initially in a superposition of the adiabatic eigen-states $|\psi
\rangle =c_{+}|\chi _{+}[x]\rangle +c_{-}|\chi _{-}[x]\rangle $ along a
certain polarization direction depending on $x$. When the particle moves so
slowly that the adiabatic condition
$$
|\frac d {dt}{x}k\sin \theta /\mu B(x)|\ll 1\eqno(4.2)
$$
holds, to the lowest order of the B-O approximation , the total initial
state $|\Psi (0)\rangle =\{c_{+}|\chi _{+}[x]\rangle +c_{-}|\chi
_{-}[x]\rangle \}\otimes |\phi (x)\rangle $ will evolves into an entangled
state
$$
|\Psi (t)\rangle =c_{+}|\chi _{+}[x]\rangle \otimes |D_{+}(t)\rangle
+c_{-}|\chi _{-}[x]\rangle \otimes |D_{-}(t)\rangle \eqno(4.3)
$$
Here, $|D_{\pm }(t)\rangle =\exp [-iH_{\pm }t]|\phi (x)\rangle $ are the
spatial states governed by the effective Hamiltonians
$$
H\pm =\frac 1{2M}(-i\partial _x-A_{\pm })^2+V_{\pm }(x)\eqno(4.4)
$$
The effective scalar potentials $V_{\pm }(x)$ and the induced vector
potentials $A_{\pm }=\frac 12k(1\pm \cos \theta )$ are determined from the
adiabatic spin eigenstates $|\chi _{+}[x]\rangle \ $and $|\chi
_{-}[x]\rangle $. In the semi-classical picture, because the particles in
the adiabatic spin states $|\chi _{+}[x]\rangle \ $and $|\chi _{-}[x]\rangle
$ separately suffer two forces $F_{\pm }=-\frac \partial {\partial x}V_{\pm
}(x)$ of opposite directions along $\mathbf{x}$, they will finally form two
macroscopically -distinguishable spots on the detecting screen, each of
which is correlated to one of the spin states. This spin-space correlation
process enables people to pick out different spin states according to the
spatial distribution.

To analyze this measurement process in details we assume the spatial part $%
\phi (x)$ in the initial state is a Gaussian wave packet
$$
|\phi (x)\rangle = \left( \frac 1{2\pi a^2}\right) ^{\frac 14}\int dx\; e^{-\frac{%
{\LARGE x}^2}{4a^2}}\;|x\rangle\eqno(4.5)
$$
distributing along direction $x$ with the center at the original point. Here
$a$ is the initial width of the atom beam. Adopting the semi-classical
method, we have the linear approximation $B(x)\simeq [\partial _xB\left(
x=0\right) ]x$ and $f=\mu \partial _xB\left( x=0\right) $. Factorizing the
evolution operator $U_{\pm }(t)=\exp [-iH_{\pm }t]$ by Wei-Norman method
[47,48](see Appendix 1), we exactly obtains,in position representation, the
following effective wave functions $|D_{\pm }(t)\rangle $at time $t$
$$
\langle x|D_{\pm }(t)\rangle =\left( \frac{a^2}{2\pi ^3}\right) ^{\frac
14}\left( \frac \pi {a^2+\frac{it}{2M}}\right) ^{\frac 12}e^{-i\Omega _{\pm
}\left( t\right) \mp iftx}\exp \left[ -\frac{\left( x-x_{\pm c}(t)\right) ^2%
}{4(a^2+\frac{it}{2M})}\right] \eqno(4.6)
$$
where
\[
\Omega _{\pm }(t)=\frac{f^2t^3}{6M}+\frac 12ft^2A_{\pm }
\]

It is seen from Eq.(4.6) that the Gaussian wave packets $\langle x|D_{\pm
}(t)\rangle $ center on the classical trajectories
$$
x_{\pm c}(t)=\mp \frac 12\cdot \frac fMt^2-\frac{A_{\pm }}Mt\eqno(4.7)
$$
They have the different group speeds $v_{\pm }=\mp \frac fMt$ $-\frac{A_{\pm
}}M$ along the opposite directions, but have the same width $a(t)=a\sqrt{1+%
t^2/(4M^2a^2)}$ spreading with time. It is obvious that the
motions of the wave packet centers obey the classical motion law that a
particle of mass $M$ forced by $\mp f$ will move with the acceleration $\mp
f/M$. The quantum character of this motion is mainly reflected in the
spreading of the wave-packets. The induced gauge fields $A_{\pm }$ are
constant, but they change the initial value of $\frac d {dt}{x}$ according
to the corresponding classical Hamilton equation.
$$
M\frac {d^2} {dt^2}{x}=\mp f;\ \ \frac d {dt}{x}=\frac pM-\frac{A_{\pm }}M
\eqno(4.8)
$$
This means that the zero initial value of the canonical momentum $p=M%
\frac d {dt}{x}+A_{\pm }$ determines the initial velocity $\frac d {dt}{x}%
(0)=-A_{\pm }/M.$ The quantum effects of $A_{\pm }$ are to contribute
the additional phases $-\frac 12ft^2A_{\pm }$in the wave functions.

The macroscopic distinguishibility of wave-packets in quantum measurement
requires that the distance between the two wave-packets should be larger
than the width of each wave packet, i.e.
$$
ft^2-k\cos \theta t\gg a\;\sqrt{M^2+\frac{t^2}{4a^2}}%
\eqno(4.9)
$$
This condition is easily satisfied for a long time evolution.

To analyze the decoherence quantitatively, we compute the norm of the
decoherence factor $F(t)=|\langle D_{+}(t)|D_{-}(t)\rangle |$. The extent of
quantum coherence depends totally on this overlapping integral. We can
explicitly integrate it
$$
F(t)=\exp \left[ -a^2f^2t^2-\frac 1{8a^2}\left(\frac fMt^2-\frac{k\cos \theta }%
Mt\right)^2\right] \eqno(4.10)
$$
It is obvious that the decoherence process indeed happens as $t\rightarrow
\infty ,$ but it does not obey the simple exponential law $e^{-\gamma t}.$
In a long time scale, the temporal behavior of decoherence is described by $%
F(t)\approx \exp \left[ -\frac{f^2t^4}{8a^2M^2}\right] $ and the
characteristic time of the decoherence process can be defined by $F(\tau
_d)=e^{-1}$ , that is
$$
\tau _d=\sqrt{\frac{2\sqrt{2}Ma}f}\eqno(4.11)
$$
This shows that the long time behavior of decoherence is indepedendent of
the spatial details of interaction , which is caused by the configuration of
the external field.

\section{Decoherence Resulting from Large Spin}

There is a second illustration to show the happening of decoherence owning
to the adiabatic separation of two systems. Based on our previous
investigation about quantum decoherence in the classical limit [17,21,22] ,
we assign an arbitrary spin $j$ to interact with a two-level system (such as
a spin-$\frac 12$ system) through a coupling of particular form. Let $%
\mathbf{J}=(\hat{J}_x,\hat{J}_y,\hat{J}_z)$ be the angular momentum operator
of the large system and $\mathbf{\sigma =(}\sigma _x,\sigma _{y,}\sigma _z%
\mathbf{)}$ be the Pauli matrix describing the quasi-spin of the two-level
quantum system with energy-level difference $\omega _s$. The full
Hamiltonian of this model is
$$
H_I=\omega _s\sigma _z\mathbf{+\omega }J\mathbf{_z+f}(\mathbf{J})\sigma _x,%
\eqno{(5.1)}
$$
The general interaction $\mathbf{f}(\mathbf{J})\sigma _x$ is linear with
respect to the variable of the quantum system while it depends on the
variable $\mathbf{J}$ through a function $\mathbf{f}(\mathbf{J})$. Two free
Hamiltonians $\omega _s\sigma _z$ and $\mathbf{\omega }J\mathbf{_z}$ were
introduced to consider the energy-exchange between the quantum system and
the large system.

The interaction $\mathbf{f}(\mathbf{J})\sigma _x$ can not well distinguish
the states $|\pm \frac 12\rangle $ of the quantum system for $|\pm \frac
12\rangle $ are not the eigen- states of the interaction Hamiltonian. So, in
general, this model can not well describe a quantum measurement process and
thus can not give a good description of quantum decoherence. However, if we
think $\mathbf{J}$ as the slowly-changing variable relative to the fast one $%
\mathbf{\sigma ,}$determined by the B-O approximation under the adiabatic
condition, the effective potential $V_{\pm }=\pm \sqrt{\omega _s^2+\mathbf{f}%
(\mathbf{J})^2}$ of the large system will clearly distinguish the adiabatic
eigen-states $|u_{+}[\mathbf{J}]\rangle =(\cos \frac \vartheta 2,\sin \frac
\vartheta 2)^T$ and $|u_{-}[\mathbf{J}]\rangle =(\sin \frac \vartheta
2,-\cos \frac \vartheta 2)^T$. Here, the angle parameter $\vartheta =\arg
\tan (-\frac{\mathbf{f}(\mathbf{J})}{\omega _s})$ depends on the slow
variable $\mathbf{J.}$ Then, the adiabatic separation of the spin-$\frac 12$
and the large spin system will result in a quantum decoherence.

In fact, because of the introduction of the arbitrary spin $j$ , which
labels the $2j+1$-dimensional irreducible representation of the rotation
group SO(3), we are able to consider the behaviors of the quantum dynamics
governed by this model Hamiltonian in the classical limit with infinite spin
$j$. The reason why the limit with infinite $j$ is called classical is that
the mean square deviations of the components $\hat{J}_x,$ and $\hat{J}_y$
enjoy the following limit feature $\frac{\Delta \hat{J}_x}j=\frac{\Delta
\hat{J}_y}j=\frac 1{\sqrt{2j}}\rightarrow 0$ as$~j\rightarrow \infty $%
[34,17,49].

To solve the dynamical evolution of the total system explicitly, we choose a
particular form of interaction : $\mathbf{f(J)}=\sqrt{g^2J_x^2-\omega _s^2}.$
Taking this particular form is equivalent to making a linear approximation
for the effective potential $V_{\pm }[\mathbf{J}].$ With this particular
form the effective Hamiltonians $H_{\pm }=\mathbf{\omega }J\mathbf{_z}+$ $%
V_{\pm }[\mathbf{J}]$ can be expressed as an rotation of the simple
spin-Hamiltonian $H_o=\sqrt{g^2+\omega ^2}J\mathbf{_z,i.e.,}$%
$$
H_{\pm }=\exp [i\hat{J}_y\phi _{\pm }]H_o\exp [-i\hat{J}_y\phi _{\pm }]%
\eqno{(5.2)}
$$
where the polar angle $\phi _{\pm }$ is defined by $\tan \phi _{\pm }=\pm
\frac g\omega $.

According to the quantum angular momentum theory, the eigen-states of $%
H_{\pm }$ can be constructed as

$$
|j,m(\phi _{\pm })\rangle =\exp [i\hat{J}_y\phi _{\pm }]|j,m\rangle
=\sum_{m=-j}^jd_{m^{\prime },m,}^j(-\phi _{\pm })|j,m^{\prime }\rangle %
\eqno{(5.3)}
$$
where $|j,m\rangle $ is a standard angular momentum state and $d_{m^{\prime
},m,}^j(\phi )=\langle j,m^{\prime }|\exp [i\hat{J}_y\phi ]|j,m\rangle $ is
the corresponding $d-$ function ; the corresponding eigen-values are $E_m=m%
\sqrt{g^2+\omega ^2}$.

Here, we should remark that the exact solvability of the above model
largely depends on the particular form of the function $\mathbf{f(J)}$. If
this is not the case, the above method can not work well and then certain
semi-classical approximation methods should be used to deal with the
effective Hamiltonian in its classical limit with very large $j$. If the
coupling function $\mathbf{f(J)}$depends on $\mathbf{J}$ quite slightly, we
can generally linearize the above effective potential $V_{\pm }(\mathbf{J})$
to realize the particular form.

We are concerned with classical characters of the large-spin system. Let us
suppose it is initially assigned the adiabatic ground state $|j,m=-j(\phi
)\rangle $ with the lowest magnetic quantum number $m=-j$. In quantum
measurement theory, the choice of ground state is required by a stable
measurement. Starting with its initial state
$$
\mid \psi (0)\rangle=(C_{+}|u_{+}[\mathbf{J}]\rangle +C_{-}|u_{-}[\mathbf{J}%
]\rangle )\otimes |j,-j(\phi )\rangle \eqno{(5.4)}
$$
the effective Hamiltonians (5.2) evolves the large spin system into an
entanglement state
$$
\mid \psi (t)\rangle=C_{+}|u_{+}[\mathbf{J}]\rangle \otimes \mid D_{+}(t)\rangle
+C_{-}|u_{-}[\mathbf{J}]\rangle \otimes \mid D_{-}(t)\rangle ,\eqno{(5.5)}
$$
with
$$
\mid D_{\pm }(t)\rangle =\exp [\pm i\hat{J}_y\phi ]\exp [-it\stackrel{\wedge
}{\mathbf{J}}\mathbf{_z}\sqrt{g^2+\omega ^2}]\exp [\mp i\hat{J}_y\phi
]|j,-j(\phi )\rangle \eqno{(5.6)}
$$

Using the explicit expressions of the $d$-function $d_{m^{\prime
},m,}^j(\phi _{\pm }),$ we can calculate the overlapping $\langle
D_{-}(t)\mid D_{+}(t)\rangle $ , obtaining
$$
F(j;t)=|\langle D_{-}(t)\mid D_{+}(t)\rangle |=\left|1-\sin ^22\phi
\sin ^2\frac{\sqrt{g^2+\omega ^2}}{2}t\right|^j.\eqno{(5.7)}
$$

The above formula directly manifests the happening of quantum decoherence in
the classical limit $j\rightarrow \infty $ . In fact, in a nontrivial case
with $\phi \neq 0$ , $|1-\sin ^2\frac t2\sqrt{g^2+\omega ^2}\sin ^22\phi |$
is usually a positive number less than 1. In the classical limit with $%
j\rightarrow \infty $ , its $j-th$ power $|\langle D_{-}(t)\mid
D_{+}(t)\rangle |$ must approach for  $t\neq t_n\equiv
2nt/\sqrt{g^2+\omega ^2},n=0,1,2...$. At those instances
$t_n,$quantum coherence revivals as so-called quantum jumps (see
Fig.3). Then, as far as the present model is concerned, we reach
the conclusion that, if the large spin system behaves classically,
the decoherence can be dynamically realized for the entangled
quantum system. In traditional quantum measurement, the detector
was pre-required as a purely classical object to reduce the
coherent superposition instantaneously. But now it is proved that
the WPC occurs as the \textit{quantum }detector moves slowly to
approach the classical limit. This means in our treatment the
detector is essentially still a quantum object. Thus it has the
advantage of dealing with the problem of quantum measurement
consistently within the framework of quantum theory.

\section{Intracavity Dynamics with Classical Source}

Our third example about decoherence in quantum adiabatic process is the
intracavity dynamics with a classical source, which is associated with the
interferometric detection of the gravitational wave by a squeezed light
[35,36].

We consider a cavity with two end mirrors (as in Fig.4), one of
which is fixed while the other is treated as a simple harmonic
oscillator of frequency $\Omega $ and mass $M$ with the position
and momentum $x$ and $p.$ The radiation pressure force of the
cavity field on the moving mirrors is proportional to the
intracavity photon density. Let $a^{\dagger }$ and $a$ be the
creation and annihilation operators of the cavity with a single
mode of frequency $\omega
$. The cavity-mirror coupling is described by an interaction Hamiltonian $%
H_I=gx$ $a^{\dagger }$ $a$ where $g$ is the coupling constant depending on
the electric dipole. In the radio frequency range the cavity field can be
prescribed as a macroscopic current. From this consideration we describe the
cavity field dynamics with the Hamiltonian $H_c=\omega a^{\dagger }$ $%
a+f(a^{\dagger }$ $+a)$. This cavity field -mirror coupling system can also
be used to detect the photon number in the cavity by the motion of mirror.
Obviously, the motion of the mirror is slow with respect to the oscillation
of the cavity field. Thus we can use the B-O approximation to approach the
quantum decoherence problem in the measurement of the cavity field. Most
recently, the special case of this model without classical source has been
used as a scheme probing the decoherence of a macroscopic object [51].

Coupled with the mirror and the classical source, the adiabatic eigen-states
$$
|n[x]\rangle =\frac 1{\sqrt{n!}}[a^{\dagger }+\lambda (x)]^n|0\rangle %
\eqno{(6.1)}
$$
of the cavity field for $\;$displacement $\lambda (x)=\frac f{\omega +gx}$
are determined by
$$
\{[\omega +gx]a^{\dagger }a+f(a^{\dagger }+a)\}|n[x]\rangle
=v_n(x)|n[x]\rangle \eqno{(6.2)}
$$
with the corresponding eigen-values $v_n(x)=n(\omega +gx),n=0,1,2,...$ Under
the B-O approximation, the effective Hamiltonians are also referred to as
the forced harmonic oscillators in the same renormalization external
potential (RNEP)$V_{rne}=\frac{f^2}{\omega +gx}$ [11,12]. Under the
adiabatic condition
$$
\left|\frac{\langle(n-1)[x]\partial _x|n[x]\rangle \frac d {dt}{x}}{\omega +gx}%
\right|\sim \frac{|ngf\frac d {dt}{x}|}{\omega ^3}\ll 1\eqno{(6.3)}
$$
$\mu ,$the RNEP $V_{rne}$ can be linearized as $\frac{f^2}\omega [1-\frac{gx}%
\omega |$ . Then the effective Hamiltonians can be rewritten as $H_n=\Omega
b^{\dagger }b+g_n(b^{\dagger }+b)$ in terms of

$$
 b=\frac{M\Omega\; x+ip}{\sqrt{2M\Omega }},\quad
 g_n=\frac{g(n-f^2/\omega^2)}{\sqrt{2M\Omega }}
    =\mu \left(n-\frac{f^2}{\omega ^2}\right)\eqno{(6.4)}
$$
For each effective Hamiltonian $H_n,$the corresponding evolution is a
displacement operator
$$
D[\alpha _n(t)]=\exp\Big(\alpha _n(t)b^{\dagger }-\alpha _n(t)^{*}b\Big)\eqno{(6.5)}
$$
with $\alpha _n(t)=-g_n(\exp [i\Omega t]-1)/\Omega.$

Let the initial state of the mirror be a well-defined quasi-classical state,
a coherent state $|\alpha \rangle $ and the initial state of the cavity be a
superposition $|c(0)\rangle =\sum_nc_n|n[x]\rangle $of the adiabatic states.
The evolution governed by the effective Hamiltonian $H_n$ leads to an
entangled state
$$
|\psi _I(t)\rangle =\sum_nc_n|n[x]\rangle \otimes D[\alpha _n(t)]|\alpha
\rangle\equiv\sum_nc_n|n[x]\rangle \otimes |D_n(t)\rangle \eqno{(6.6)}
$$
for the total system. The overlapping of the mirror states in this
entanglement can be computed and its norm is
$$
|\langle D_m(t)|D_n(t)\rangle |=\exp\left(-(n-m)^2\frac{2\mu ^2}{\Omega ^2}\sin
^2\frac{\Omega t}2\right)\eqno{(6.7)}
$$

The changing rate $\frac d {dt}{x}($ the velocity ) of the slow variable $x$
is proportional to $\Omega .$ In the adiabatic limit, $\Omega $ is very
small. So we can rationally consider the limit $\Omega \rightarrow 0$ for a
fixed $\mu $. Then an ideal entanglement appears in this limit case for the
overlapping becomes an non-linear exponential decaying factor
$$
|\langle D_m(t)|D_n(t)\rangle |=\exp\left(-\frac12(n-m)^2\mu ^2t^2\right)\eqno{(6.8)}
$$
This result is quite similar to that of the Cini model in van Hove limit
[34]. This decay phenomenon was first illustrated in ref.[21,23].
Mathematically,it results from the fact that in the strong coupling
limit, the period of the oscillation is very large in comparison with the
small frequency $\Omega$.

Another interesting situation arise when the mirror is initially prepared in
a Fock state $|n\rangle =\frac 1{\sqrt{n}!}\left( a^{\dagger }\right)
^n|0\rangle $. To show a macroscopic, but non-classical dynamic behavior,
the Fock state should possess a very large occupation number $n$. The
overlapping for the initial Fock state can be expressed as
$$
F(t,n)=\langle n|D[-\alpha _k(t)]D[\alpha _l(t)]|n\rangle
$$
$$
=\exp [-\frac 12(l-k)^2\frac{\mu ^2}{\Omega ^2}\sin ^2\frac{\Omega t}%
2]L_n\left( (l-k)^2\frac{\mu ^2}{\Omega ^2}\sin ^2\frac{\Omega
t}2\right) \eqno{(6.9)} $$ in terms of the Laguerre polynomial
$L_n(z)$.  Fig.5 shows  $F(t,n)$ as a function of time $t$ for
different $j$. In fact, according to the theory of
special function, $L_n(z)$ approaches the zero-order Bessel function $J_0(%
\sqrt{n}z)$ when $n\rightarrow \infty $, hence [52],
$$
F(t,n)\rightarrow e^{-\frac 12(l-k)^2\mu ^2t^2}\;L_n((l-k)^2\mu
^2t^2/4)\rightarrow e^{-\frac 12(l-k)^2\mu ^2t^2}\;J_0(\sqrt{n}(l-k)^2\mu ^2t^2)%
\eqno{(6.10)}
$$
The zero-order Bessel function of real variable $\zeta \sqrt{n}$ is a
decaying-oscillating function and approaches zero as $n$ tends to infinity.
Therefore, when the cavity is occupied by a large number of photons, the
macroscopic feature of the detector (the end mirror ) dynamically decoheres
the initial pure state of the cavity.

\section{Localization of Macroscopic Object Through Adiabatic Entanglement}

As another interesting application of the above adiabatic approach for
decohernce, we will discuss how the adiabatic entanglement result in the
spatial localization of a macroscopic object. This discussion is devoted to
consider the quantum decoherence of the slow part rather than that of the
fast part, which has been studied in previous sections.

The localization problem originated from the correspondence between Einstein
and Born [53] and is closely related to the Schrodinger cat. They observed
that, usually in a spatially-localized state, a macroscopic object can only
be described by a time-dependent localized wave packet, which is a coherent
superposition of the eigen-states of the center-of-mass Hamiltonian $%
H_0=p^2/2M$ . Though the spreading of an initially well localized wave packet
can be reasonably ignored for the macroscopic object with very large mass,
Einstein argued that the superposition of two narrow wave packets is no
longer narrow with respect to the macro-coordinate, but it is still a
possible state of the macroscopic object. So a contradiction to the
superposition principle arises because of the requirement that the wave
function of a macroscopic object must be narrow. To solve this problem,
Wigner [54], Joos and Zeh [55] present the so called scattering -induced
-decoherence mechanism (or WJZ mechanism): scattering of photons or atoms
off a macroscopic object records the information of its position to form a
quantum measurement about the position. In this mechanism the interference
terms between different positions are destroyed by the generalized
''which-way''detection . In spirit of Omnes 's observation [56], we argue
that, mentioning macroscopicness implies the requirement that the
macroscopic object must contain a large number of internal blocks. Then the
macroscopic object,coupled to the internal variables, should be described by
collective variables subject to the interaction similar to that concerning
the external scattering in WJZ mechanism. In this section we will show that
the spatial localization of a macroscopic object can be caused by an ideal
entanglement between its collective position (or center-of-mass ) and
internal variables . This entanglement results from their adiabatic
separation..

Let $x$ and $q$ be, respectively, the collective position and internal
variables of a macroscopic object with the collective Hamiltonian $H_s=p^2/2M$
($[x,p]=i$). To consider how different positions affect the quantum
coherence of the internal motion of the macroscopic object, we suppose that
the total system is initially in a product state
$$
|\Psi _x(t=0)\rangle =|x\rangle \otimes |\phi \rangle \eqno{(7.1)}
$$
where the first component $|x\rangle $ is the eigen-state of the collective
position operator $x$ while $|\phi \rangle $ is an arbitrary initial pure
state of the internal degrees of freedom. Usually, the collective motion
acts on the internal motion in certain ways and the back-action of the
internal motion can not be neglected physically. So this generic interaction
can not produce an ideal entanglement between the collective position and
the internal states of the macroscopic object. By an argument similar to
that by Joos and Zeh [54] , who deal with quantum decoherence and its
relevant localization by considering the scattering of external particles by
the macroscopic object, we see only when the back-action is negligiblly
small, can the interaction between the collective and internal states
realize a ''measurement-like process'':
$$
|x\rangle \otimes |\phi \rangle \rightarrow U(t)|x\rangle \otimes |\phi
\rangle =|x(t)\rangle \otimes S(x;t)|\phi \rangle \eqno{(7.2)}
$$
Here, $U(t)$ is the total evolution matrix and $|x(t)\rangle $ $%
=U_0(t)|x\rangle $ represents the free evolution in the absence of the
coupling to the internal variables; $S(x,t)$, acting on the internal states,
denotes the effective $S-matrix$ parametrized by the collective position $x$%
. If the collective motion is initially described by a wave packet $|\varphi
\rangle =\int \varphi (x)|x\rangle dx,$then the reduced density matrix of
the collective motion is
$$
\rho (x,x^{\prime },t)=\varphi (x,t)\varphi^*(x^{\prime },t)\langle \phi
|S^{\dagger }(x^{\prime };t)S(x;t)|\phi \rangle \eqno{(7.3)}
$$
Considering the translational invariance of the scattering process, Joos and
Zeh showed that, in $x-representation$, the off-diagonal terms take the
following form
$$
\langle \phi |S^{\dagger }(x^{\prime };t)S(x;t)|\phi \rangle \sim \exp
(-\Lambda t|x-x^{\prime }|^2)\eqno{(7.4)}
$$
This means the decoherence factor is a damping function with the
localization rate $\Lambda $ , which depends on the total cross section.

Now, the question arises whether the negligibility of the back-action is the
unique cause for the appearance of the above mentioned ''measurement-like
process''. If not, what are the other causes beyond it? To resolve this
problem, we assume the Hamiltonian $h(q,x)=H_i(q)+W(x,q)$ describes the
motion of the internal variables $q$ coupling to the collective variable $x$%
. For a fixed value of the slow variable $x$, the eigen-state $|n[x]\rangle $
and the corresponding eigen-values $V_n[x]$ are determined by the eigen-
equation $h(q,x)|n[x]\rangle =V_n(x)|n[x]\rangle $. Regarding $x$ and $q$ as
the slow and fast variables respectively in the BO adiabatic approach , we
approximately obtain the complete set \{$\phi _{n,\alpha }(x)|n[x]\rangle $%
\} of eigenstates of the total system, where $\phi _{n,\alpha }(x)$ come
from the eigen-equation $H_n\phi _{n,\alpha }(x)=E_{n,\alpha }\phi
_{n,\alpha }(x)$ and $H_n=p^2/M+V_n[x]$ is the effective Hamiltonian
correlated to the internal state $|n[x]\rangle $. Here, we do not consider
the induce gauge potential. Then,we can see how the ``measurement-like
process'' naturally appears as a result of the adiabatic dynamic evolution.

In fact, under the BO approximation, we can expand the factorized initial
state $|\Psi (0)$ $\rangle =$ $|x\rangle \otimes |\phi \rangle $ in terms of
the adiabatic basis \{$\langle x|n,\alpha \rangle \equiv \phi _{n,\alpha
}(x)|n[x]\rangle $\} and then we obtain the total wave function
\[
|\Psi (t)\rangle =\sum_{n,\alpha }\langle \phi _{n,\alpha }|x\rangle \langle
n[x]|\phi \rangle e^{-iE_{n,\alpha }t}|n,\alpha \rangle
\]
$$
=\sum_n\langle n[x]|\phi \rangle \int dx^{\prime }\langle x^{\prime
}|e^{-iH_nt}|x\rangle |x^{\prime }\rangle \otimes |n[x^{\prime }]\rangle %
\eqno{(7.5)}
$$
where we have used the single-component completeness relation $\sum_\alpha
|\phi _{n,\alpha }\rangle \langle \phi _{n,\alpha }|=1.$Generally, the
propagator $K(x^{\prime },x,t)=\langle x^{\prime }|e^{-iH_nt}|x\rangle
|x^{\prime }\rangle $ is not diagonal for $|x\rangle $ is not an eigen-state
of $H_n.$ However, in the large mass limit , we can prove that, to the first
order approximation , $K(x^{\prime },x,t)$ takes a diagonal form
proportional to a $\delta -function.$ In fact,in this case, the kinetic term
$p^2/2M$ can be regarded as a perturbation in comparison with the effective
potential $V_n(x).$ Using Dyson expansion to the first order of $\frac 1M$,
we have
\[
e^{-iH_nt}=e^{-iV_nt}\left(1-i\int_0^te^{iV_nt^{\prime }}\frac{p^2}{2M}%
e^{-iV_nt^{\prime }}dt^{\prime }+\cdots\right)
\]
$$
=e^{-iV_nt}\left(1-i\frac{p^2t^2}{2M}+i\frac{t^2}{4M}(p\partial _xV_n+[\partial
_xV_n]p)-\frac{it^3\partial _xV_n^2}{6M}+cdots\right)\eqno{(7.6)}
$$
Since $\int \langle x^{\prime }|P^n|x\rangle f(x^{\prime })dx=0$ for
n=1,2,..., we have
$$
K(x^{\prime },x,t)=e^{-iV_n[x]t}[\delta (x-x^{\prime })+\frac
i{2M}\int_0^td\tau e^{-iV_n[x^{\prime }]\tau }\frac{\partial ^2}{\partial
x^{\prime 2}}\delta (x-x^{\prime })e^{iV_n(x)\tau }]\eqno{(7.7)}
$$
We notice this simple result has the following physical explanation: the
evolution state of a heavy particle for very large $M$, which is almost
steady, is approximately an eigenstate of the position operator if it is
initially in a state with a fixed position. Then,it follows that, in the
large-mass limit, the wave function $|\Psi (t)\rangle $ can be factorized
approximately: $|\Psi (t)\rangle =|x\rangle \otimes S(x,t)|\phi \rangle $
where the entangling $S-matrices$%
$$
S(x,t)=\sum_{n,}e^{-ihV_nt}|n[x]\rangle \langle n[x]|\eqno{(7.10)}
$$
are defined in terms of the adiabatic projection $|n[x]\rangle \langle n[x]|$%
.

According to our previous argument about the factorized structure of $%
S-matrix$ in the dynamic theory of quantum measurement [ ], if the internal
degree of freedom has many components, e.g.,if $q=(q_1,q_2,...q_N)$ ,then in
their normal non-interaction modes , $S(x;t)$ can be factorized
as:
$$
S(x;t)=\prod_{j=1}^NS_{_j}(x;t)\eqno{(7.11)}
$$
with
$$
S_{_j}(x;t)=e^{-ih_j(q_j,x)t}\eqno{(7.12)}
$$
with $h(q,x)=\sum_jh_j(q_j,x)$. Of course in the derivation of the above
factorized structure for the $S-matrix$ , we have made some simplifications.
Roughly speaking, we have assumed that the potential takes the form of
direct sum and the eigenstate the form of direct product

$$
V_n=\sum_jV_{nj}(q_j),\qquad |n[x]\rangle =\prod_{j=1}^N\otimes
|n_j[x]\rangle \eqno{(7.13)}
$$
neglecting the higher order terms $\approx $ $O(\frac 1M).$

For the initial state $|\phi \rangle =\prod_{j=1}^N\otimes $ $|\phi
_j\rangle $ factorized with respect to internal components, the reduced
density matrix
$$
\rho (x,x^{\prime },t)=\varphi (x)\varphi^*(x^{\prime })F_N(x^{\prime },x,t):%
\eqno{(7.14)}
$$
can be re-written in terms of the so called decoherence factor
$$
F_N(x^{\prime },x,t)=\prod_{j=1}^NF^{[j]}(x^{\prime },x,t)\equiv
\prod_{j=1}^N\langle \phi _j|S_{q_j}^{\dagger }(x^{\prime
};t)S_{q_j}(x;t)|\phi _j\rangle .\eqno{(7.15)}
$$
This factor is expressed as an $N$-multiple product of the single decohering
factors $F^j(x,x^{\prime })=$ $\langle \phi _j|S_{q_j}^{\dagger }(x^{\prime
};t)S_{q_j}(x;t)|\phi _j\rangle $with norms less than unity. Thus in the
macroscopic limit $N\rightarrow \infty $ , it is possible that $%
F_N(x^{\prime },x,t)$ $\rightarrow 0,$ for $x^{\prime }\neq x$. In fact,
this factor reflects almost all the dynamic features of the influence of the
fast part on the slow part. Physically, an infinite $N$ means that the
object is macroscopic since it is made of infinite number of particles in
that case. On the other hand, the happening of decoherence at infinite $N$
manifests a transition of the object from the quantum realm to the classical
realm.Here,as expected,the physical picture is consistent.

As to the localization problem raised by Einstein and Born [53], we , based
on the above argument, comment that one can formally write down the wave
function of a macroscopic object as an narrow pure state wave packet, but it
is not the whole of a real story. Actually, the statement that an object is
macroscopic should physically imply that it contains many particles. So a
physically correct description of its state must concern its internal
motions coupling to the collective coordinates (e.g., its center-of-mass) .
Usually, one observe this collective coordinate to determine whether two
spatially-localized wave packets can interfere with each other. If there
does not exist such interference, one may say that, the superposition of two
narrow wave packets for the macro-coordinate is no longer a possible pure
state of the macroscopic object. Indeed, because the ``which-way''
information of the macro-coordinate is recorded by the internal motions of
particles making up the macroscopic object, the induced decohernce must
destruct the coherence in the original superposition so that the state of
the macroscopic object is no longer pure.

The present argument also provides a possible solution for the Schroedinger
cat paradox. If we consider the Schroedinger cat as a macroscopic object
consisting of many internal particles, then we can never observe anything
corresponding to the interference between the dead and the living cats
because the macroscopically- dead and the macroscopically- living states of
the cat are correlated to the corresponding internal states. In this sense,
we conclude that the Schroedinger cat paradox is not a paradox at all in
practice. Rather, it essentially arises from overlooking the internal
motions of a macroscopic cat or the multi-particle scattering off it.
Turning to the problem of quantum coherence of a subsytem within the total
system, from the above argument we also conclude that the classical
characters of both the quantum system and the large system entangled with it
are ``correlated'' physically: when the large system transits from quantum
to classical realms, the quantum system has to act in the same way. In other
words, you can never see a coherent superposition of microscopic states
entangled to a live or a dead cat's states if the Schrodinger cat is
classical. In the presence of a classical cat, the quantum system entangling
with it should lose its own coherence. Actually, if , in a classical
manner,one asks experimentally what a quantum system really does , then the
quantum system would behave physically like a classical object . This is
just the quantum mystery physicists have to face.

To make a deeper elucidation of the above general arguments about the
localization of a macroscopic object of mass $M$, we model the macroscopic
object as consisting of $N$ two level particles, which are fixed at certain
positions to form a whole without internal spatial motion. The collective
position $x$ is taken to be its mass-center or any reference position on it
while the internal variables are the quasi-spins associated with two level
particles. Generally, if we assume the back-action of the internal variables
on the collective position is relatively small, the model Hamiltonian can be
written as

$$
H=\frac{P^2}{2M}+\sum_{j=1}^N[f_j(x)|e_j\rangle \langle
g_j|+f_j^*(x)|g_j\rangle \langle e_j|]+\sum_{j=1}^N\omega _j[|e_j\rangle
\langle e_j|-|g_j\rangle \langle g_j|\\[-4mm]
]\eqno{(7.16)}
$$
where $|g_j\rangle $ and $|e_j\rangle $ are the ground and the excited
states of the $j$ 'th particle and $f_j(x)$ denote the position-dependent
couplings of the collective variable to the internal variables. Let $l_j$ be
the relative distance between the $j$ 'th particle and the reference
position $x.$We can further assume $f_j(x)=f(x+l_j).$ Physically,we may
think that these couplings are induced by an inhomogeneous external field ,
e.g., they may be the electric dipole couplings of two-level atoms in an
inhomogeneous electric field.

We remark that the above model enjoys some universality under certain
conditions, compared with various environment models inducing both
dissipation and decoherence of quantum processes. In fact, Caldeira and
Leggett [25] have pointed out that any environment weakly coupling to a
system may be approximated by a bath of oscillators under the condition that
``each environmental degree of freedom is only weakly perturbed by its
interaction with the system''. We observe that any linear coupling only
involves transitions between the lowest two levels (ground state and the
first excitation state) of each harmonic oscillator in the perturbation
approach though it has many energy levels. Therefore in such a case we can
also describe the environment as a combination of many two level subsystems
without losing generality [28].To some extent, these arguments justify our
choosing the two level subsystems to model the internal motion of the
macroscopic object.We will soon see its advantage:the localization
characters can be manifested naturally and clearly.

Now let us calculate the $S_{_j}(x;t)$ for this concrete model. The
single-particle Hamiltonian $h_j(x)=\omega _j(|e_j\rangle \langle
e_j|-|g_j\rangle \langle g_j|)+(f_j(x)|e_j\rangle \langle g_j|+h.c)$ has the
$x$-dependent eigenvalues
$$
V_{jc}=n\Omega _j(x)\equiv \pm \sqrt{|f_j(x)|^2+\omega _j^2}\qquad (n=\pm) %
\eqno{(7.17)}
$$
and the corresponding eigen-vectors$|n_j[x]\rangle $ are
$$
|+_j[x]\rangle =\cos \frac{\theta _j}2|e_j\rangle +\sin \frac{\theta _j}%
2|g_j\rangle,\eqno{(7.18)}
$$
$$
|-_j[x]\rangle =\sin \frac{\theta _j}2|e_j\rangle -\cos \frac{\theta _j}%
2|g_j\rangle, \eqno{(7.19)}
$$
where $\tan \theta _j=\frac{f_j(x)}{\omega _j}.$ Then we explicitly the
corresponding single-particle $S-matrix$

$$
S_{_j}(x;t)=\left(
\begin{array}{cc}
\cos (\Omega _jt)-i\sin (\Omega _jt)\cos \theta _j, & i\sin (\Omega _jt)\sin
\theta _j \\
i\sin (\Omega _jt)\sin \theta _j, & \cos (\Omega _jt)+i\sin (\Omega _jt)\cos
\theta _j
\end{array}
\right) \eqno{(7.20)}
$$
Here in the derivation we have used the formula $\exp [i\overrightarrow{%
\sigma }\cdot \overrightarrow{A}]=\cos A+i\overrightarrow{\sigma }\cdot
\overrightarrow{n_A}\sin A$ for a given vector $\overrightarrow{A}$ of norm $%
A$ along the direction $\overrightarrow{n_A}.$Having obtained the above
analytic results about $S-matrix,$ we can further calculate the
single-particle decoherence factors $F^{[j]}(x^{\prime },x,t)\equiv \langle
g_j|S_{_j}^{\dagger }(x^{\prime };t)S_{_j}(x;t)|g_j\rangle $for a given
initial state $|\phi \rangle =\prod_{j=1}^N\otimes $ $|g_j\rangle $. For
simplicity we use the notation $f(x^{\prime })=f^{\prime }$ .We have
\[
F^{[j]}(x^{\prime },x,t)=\{\sin (\Omega _j^{\prime }t)\sin \theta _j^{\prime
}\sin (\Omega _jt)\sin \theta _j+
\]
$$
\cos (\Omega _j^{\prime }t)\cos (\Omega _jt)+\sin (\Omega _j^{\prime }t)\cos
\theta _j^{\prime }\sin (\Omega _jt)\cos \theta _j^{\prime }\cos \theta _j%
\eqno{(7.22)}
$$
\[
+i\{\cos (\Omega _j^{\prime }t)\sin (\Omega _jt)\cos \theta _j-\sin (\Omega
_j^{\prime }t)\cos \theta _j^{\prime }\cos (\Omega _jt)\}\}
\]
In the weakly coupling limit with $g_j\ll \omega _j$ and the coupling $%
f_j\simeq g_jx$,we have $\sin \theta _j\simeq \theta _j\simeq \frac{f_j}{%
\omega _j},\cos \theta _j\simeq 1-\frac 12\theta _j^2$ and $\Omega _j\simeq
\omega _j.$Thus, the decohering factors can be simplified as
$$
F^{[j]}(x^{\prime },x,t)\simeq 1-(x-x^{\prime })^2\frac{|g_j|^2}{2\omega _j^2%
}\sin ^2(\omega _jt)\\[-4mm]
+\frac{i|g_j|^2}{4\omega _j^2}\{x^2-x^{\prime 2})\sin (2\omega _jt)%
\eqno{(7.23)}
$$
Consequently, the temporal behavior of the decoherence is determined by
$$
F(x^{\prime },x,t)=\exp\left\{-(x-x^{\prime })^2\sum_{j=1}^N\frac{|g_j|^2}{%
2\omega _j^2}\sin ^2(\omega _jt)+\\[-4mm]
(x^2-x^{\prime 2}{})\sum_{j=1}^N\frac{i|g_j|^2}{4\omega _j^2}\sin (2\omega
_jt)\right\}\eqno{(7.24)}
$$

In the case of continuous spectrum, the sum $R(t)=$ $\sum\limits_{j=1}^N%
\frac{g_j^2}{2\omega _j^2}\sin ^2\left( \omega _jt\right) $can be re-
expressed in terms of a spectrum distribution $\rho (\omega _k)$ as $%
R(t)=\int_0^\infty \frac{\rho (\omega _k)g_k^2}{2\omega _k^2}\sin ^2\omega
_kd\omega _k.$From some concrete spectrum distributions, interesting
circumstances may arise. For instance, when $\rho (\omega _k)=\frac{4}\pi
\gamma /g_k^2$ the integral converges to a negative number proportional to
time t , precisely, $S(t)=\gamma t$ [31] .

Therefore, our analysis recovers the result
$$
\rho (x,x^{\prime },t)=\varphi (x)\varphi^*(x^{\prime })e^{-\gamma
t(x-x^{\prime })^2}\exp [i\pi (x^2-x^{\prime 2}{})s(t)]\eqno{(7.25)}
$$
for the reduced density matrix of the macroscopic object , which was
obtained by Joos and Zeh [54] through the multi particle external scattering
mechanism and by Zurek separately through Markov master equation. Here, $%
s(t)=\sum_{j=1}^N\frac{\sin (2\omega _jt)}{\pi \omega _j^2}$ is a
time-dependent multi-period function. This shows that the norm of the
decoherence factor is exponentially decaying and as $t\rightarrow \infty ,$
the off-diagonal elements of the density matrix vanish simultaneously! Due
to the presence of the oscillating factor $s(t)$ of multi-period, $\rho
(x,x^{\prime },t)$ seems very complicated.But on the other hand, the simple
decaying norm of $\rho (x,x^{\prime },t)$ can well serve to describe
decoherence of the macroscopic object. Consider now a similar example by
Joos and Zeh .We take a coherent superposition of two Gaussian wave packets
of width $d$
$$
\varphi (x)=\frac 1{\sqrt[4]{8\pi d^2}}\left\{\exp\left(-\frac{(x-a)^2}{4d^2}%
\right)+\exp\left(-\frac{(x+a)^2}{4d^2}\right)\right\}\eqno{(7.26)}
$$
The norm of the corresponding reduced density matrix
$$
|\rho (x,x^{\prime },t)|=\sum_{k,l=0}^1P_{kl}(x,x^{\prime },t)\eqno{(7.27)}
$$
contains 4 peaks
\[
P_{11}(x,x^{\prime },t)=\frac 1{\sqrt{8\pi d^2}}e^{-\gamma
t(x-x^{\prime })^2}\exp [-\frac{(x-a)^2}{4d^2}-\frac{(x^{\prime }-a)^2}{4d^2}%
]
\]
\[
P_{10}(x,x^{\prime },t)=\frac 1{\sqrt{8\pi d^2}}e^{-\gamma
t(x-x^{\prime })^2}\exp [-\frac{(x-a)^2}{4d^2}-\frac{(x^{\prime }+a)^2}{4d^2}%
]
\]
$$
P_{01}(x,x^{\prime },t)=\frac 1{\sqrt{8\pi d^2}}e^{-\gamma
t(x-x^{\prime })^2}\exp [-\frac{(x+a)^2}{4d^2}-\frac{(x^{\prime }-a)^2}{4d^2}%
]\eqno{(7.28)}
$$
\[
P_{00}(x,x^{\prime },t)=\frac 1{\sqrt{8\pi d^2}}e^{-\gamma
t(x-x^{\prime })^2}\exp [-\frac{(x+a)^2}{4d^2}-\frac{(x^{\prime }+a)^2}{4d^2}%
]
\]
centering respectively around the points $(a,a),(a,-a),(-a,a)$ and
$(-a,-a)$ on $x-x^{\prime }$-plane. The heights are respectively
$1/\sqrt{8\pi d^2}, e^{-4\gamma ta^2}/\sqrt{8\pi d^2},e^{-4\gamma
ta^2}/\sqrt {8\pi d^2}$and $1/\sqrt{8\pi d^2})$. Obviously, two
peaks with centers at $(a,-a)$ and $(a,-a)$ decays with time while
the other two keep their heights constant . Fig.6. shows
this time-dependent configuration at t=0,and a finite t. As $t\rightarrow \infty $%
, two off-diagonal terms $P_{10}$ and $P_{01}$decay to zero so that the
interference of the two Gaussian wave packets are destroyed . In this sense,
we say that the pure state $\rho (x,x^{\prime },t=0)=\int dx\varphi
(x)\varphi *(x^{\prime })|x\rangle \langle x^{\prime }|$ becomes a mixture
$$
\rho (t)=\int dx\varphi (x)\varphi^*(x)|x\rangle \langle x|\eqno{(7.29)}
$$
in $x-$representation.\qquad

Interference of two plane waves of wave vector $k_1,k_2$ provides us another
simplest example. Without decoherence induced by its internal motions or the
external scattering , their coherent superposition $\varphi (x)=\sqrt{\frac
1{4\pi }}[e^{ik_1x}+e^{ik_2x}]$ yields a spatial interference described by
the reduced density matrix
\[
\rho _0(x,x^{\prime },t)=\frac 1{4\pi }\{e^{ik_1(x-x^{\prime
})}+e^{ik_2(x-x^{\prime })}+
\]
$$
\exp [i(\frac{k_1^2t-k_2^2t}{2m}+k_2x-k_1x^{\prime})]+\exp [i(\frac{%
k_2^2t-k_1^2t}{2m}+k_1x-k_2x^{\prime }]\}\eqno{(7.30)}
$$
Under the influence of internal motions , it becomes
\[
\rho (x,x^{\prime },t)\approx \rho _0(x,x^{\prime },t)e^{-\gamma
t(x-x^{\prime })^2}
\]
for large mass. We see that the difference created by decoherence is only
reflected in the off-diagonal elements,and the pure decoherence (without
dissipation) does not destroy the interference pattern described by the
diagonal term
$$
\rho (x,x,t)=\rho _0(x,x,t)=\frac 1{2\pi }\{1+\cos [\frac{k_1^2t-k_2^2t}{2m}%
+(k_2-k_1)x]\}\eqno{(7.32)}
$$
This simple illustration tells us that the present quantum decoherence
mechanism may not have to do with the interference pattern of the first
order coherence, but it does destroy the higher order quantum coherence: $%
\rho (x,x^{\prime },t)\rightarrow 0$ as $t\rightarrow \infty .$ In fact,
Savage, Walls and Yu have shown that, due to the induced loss of energy,
quantum dissipation is responsible for the disappearance of the interference
pattern of the first order coherence . The influences of internal motions or
external scattering on the decoherence of a macroscopic object may be very
complicated. Intuitively, these dynamic effects should depend on the details
of interaction between the collective variables and the internal and
external degrees of freedom. Pratically,we can classify these influences
into two species,namely, quantum dissipation and quantum decoherence, and
then study them separately by different models.

\quad

\section{Concluding Remarks}

We remark that the quantum decoherence of a small system, resulting from a
transition of the entangling ``large system'' from quantum to classical, is
certainly an irreversible process. This is because the density matrices $%
\rho _s(0)$ and $\rho _s(t)$ have different ranks for $t\neq 0$. Thus they
can not be transformed into each other through an unitary time-evolution
matrix. If we only consider a closed system, as a postulate with certain
classical elements in it, its WPC or quantum decoherence can not be derived
from Schr\"{o}edinger equation based on the basic laws of quantum mechanics.
Since quantum mechanics was founded, physicists have wished to add this WPC
postulate to the axiom system of quantum mechanics. von Neumann and Wigner
made the first attempt and considered the measurement detector plus the
measured system as a total system called a``universe'' satisfying
Schr\"{o}edinger equation. They hoped that, projected on the system, the
evolution of the ``universe'' leads to wave packet collapse naturally.
However,because it did not take the macroscopic or classical character of
detector into account [1,2], this approach brings philosophical difficulty:
If the observation of the detector force the measured system to decohere,
the detector must decohere in advance. So the second detector is needed to
monitor the first one, and the third one is needed to monitor the second one
and so on. By this argument in logic, a chain of detectors should be
introduced in sequence (we usually call it von Neumann 's chain),and at the
end of this von Neumann 's chain, there should exist a pair of eyes as a
special detector ,which is required to be classical.

In 1972, to avoid the introduction of this chain of detectors, Hepp and
Coleman raised a dynamical description for the WPC via a simple
exactly-solvable model. They emphasized that, if the macroscopic limit of
the first detector is considered appropriately, detectors other than the
first one are not necessary. Using the macroscopic character of the first
detector is crucial to the solution of this problem. Later Namiki, Nakazato
and Pascazio et al generalized this work to put forward various new models
for quantum measurement[26,27]. In 1993, after carefully analyzing these
models and taking the classical limit of detector into account, one of the
authors (CPS) found that the essence implied by these models is a
factorization structure [17,18]. By exact- solvable models and
approximately-solvable models as well,it is shown that when the effective
evolution of the detector can be factorized,in the macroscopic limit that
the number of particles composing the detector approachs infinity, quantum
decoherence or WPC will appear naturally.

Previously we also associated quantum decoherence problem with the
requirement that the result of a measurement should be macroscopically
observable so that an ideal entanglement happens dynamically [21]. But all
discussions about the interaction induced quantum decoherence strongly rely
on the particular forms of interaction, namely, the interaction [3] $%
H_I(q,x) $ of non-demolition type, which depends on the variable $q$ of the
measured system, but commutes with its free Hamiltonian. Therefore, a fatal
defect inherent in the previous works,including our own works,concerning the
study of environment induced decoherence based on quantum measurement theory
is that the question why should nature choose such a particular form of
interaction remains unanswered. In some sense, the present work in this
paper has well tackled this problem.Indeed, through the B-O adiabatic
separation of the quantum and quasi-classical variables,we have demonstrated
that in the adiabatic limit, the effective interaction reduced from a quite
general coupling just takes such a particular form.

In a wide sense, the adiabatic entanglement can be well understood in the
picture of coupled channels [50], which is an extensive generalization of
B-O approximation. Consider a total system whose wave function depends on
two set of variables, $q$ and $x$. Let $Q$ be an operator only acting on the
function of $q$ and has a complete set of eigen-vectors $\{|n\rangle \}$
with the corresponding eigenvalues $v_n.$ Since $\{|n\rangle \}$ forms a
complete basis of the Hilbert space of all functions of $q,$ the total
eigenfunction $\Psi _E(x,q)$ of the full Hamiltonian $%
H=H_E(x)+H_s(q)+H_I(x,q)$ with eigen-value $E$ can be regarded as a function
of $q$ for a given $x$ and then can be expressed as $\Psi _E(x,q)=\sum \phi
_n(x)|n\rangle $. The \textit{channel wave function }$\phi _n(x)$ is defined
by the \textit{coupled channel equations}
$$
H_{nn}(x)\phi _n(x)+\sum_{m\neq n}H_{nm}(x)\phi _m(x)=E\phi _n(x)\eqno(8.1)
$$
The matrix elements $H_{mn}(x)=\langle m|H|n\rangle _q$ are defined in terms
of the $q$-function space ``integral''.Under a certain condition, if the
off-diagonal elements \textit{can be neglected physically,} an effective
non-demolition Hamiltonian $H_{eff}=H_{E-eff}(x)+H_{s-eff}+H_{in}(x)$:

\[
H_{E-eff}=diag.[H_{11}^E(x),H_{22}^E(x),.....,H_{dd}^E(x)]
\]
$$
H_{s-eff}=diag.[\lambda _1,\lambda _2,....,\lambda _d],\eqno(8.2)
$$
\[
H_{in}(x)=diag.[H_{11}^s(x),H_{22}^s(x),.....,H_d^s(x)]
\]
can be partially diagonalized in the `channel space'. Here $%
H_{mm}^A(x)=\langle m|H_A|m\rangle _q$ for $A=E,S$ and $\lambda _m=\langle
m|H_s(q)|m\rangle _q$ \textit{are constants}. Obviously, the non-demolition
condition [$H_{s-eff},$ $H_{in}(x)]=0$ holds as $H_{s-eff}$ is a constant
matrix. In the B-O approximation the channel operator $Q$ is taken to be $%
Q[x]=H_s(q)+H_I(x,q),$which is parametrized by $x.$ The adiabatic condition
maintains that, only the diagonal elements play a dominant role and the
off-diagonal elements \textit{can be neglected for very small
channel-channel coupling[42] . }Therefore, it can be concluded that there
may exist a more universal mechanism beyond B-O approximation to realize the
quantum decoherence dynamically originated from the basic interaction ,
which is related to the theory of coupled channels.\textit{\ }

Finally we point out that the presence of non-demolition interaction [3] is
only a necessary condition for quantum decoherence to appear.Sufficient
conditions should include the requirement that the large system be classical
so that its final states could be orthogonal to one another. In this
paper,we have regarded the spin-system with a very large spin and the
harmonic oscillator initially in a coherent state as classical objects. Then
within the semi-classical framework,even in the case of a general potential
motion, we are able to relate the macroscopic distinguishibility of the
quantum states of the large system to its classical limit behaviors.
However, there are still vague points in the definition of the
quantum-classical division for the large system. This problem is deeply
rooted in the following more fundamental and more challenging issue: why or
in what sense does a general large system behave classically. If we imagine
that, beside the considered quantum system, there is another system coupling
with the large system to decohere it, then the present problem will be
trapped into an evil logic chain. One notices the difficulty here is very
similar to that faced by von Neumann and Wigner about sixty years ago [1,2].
Though new experiments have been revitalizing the study of decoherence
problem and progress is being made,it seems that there is still a long way
to go to finally understand quantum irreversible process completely.To reach
this goal,one should first find a satisfactory definition for the so called
quantum-classical boundary.At present it is very unclear to us how to do
this without recourse to particular physical systems.

\section*{Acknowlegement}

\noindent \textit{This work is supported by direct grant (Project
ID:2060150) from The Chinese University of Hong Kong. It is also
partially supported by the NFS of China. One of the authors (CPS)}
\textit{wishs to express his sincere thanks to P.T.Leung, C.K.Law
and K.Young for many useful discussions. }

\newpage

\centerline{\textbf{Appendix.
Wei-Norman Algebraic Solution For $H=\frac 1{2M}(%
\mathbf{p}-A)^2+fx$}}
\vskip 0.5cm

\ Let $U(t)$ be the evolution operator of a quantum system with the
effective Hamiltonian
$$
H=\frac 1{2m}(\mathbf{p}-A)^2+f\mathbf{x}\qquad \eqno(a1)
$$
where $A$ is a constant induced gauge potential . Neglecting the constant
term, we can rewrite the effective Hamiltonian
$$
H=\frac 1{2m}\mathbf{p}^2-\frac AM\mathbf{p}+fx\eqno(a2)
$$
as an element of the Lie algebra {\Large \pounds\ }generated by $%
\{p^2,p,x,1\}.$

According to Wei-Noman's algebraic theorem [46], a solution of the
Schroedinger equation for the evolution operator $U(t)$ must be an element
belonging to the Lie group related to the Lie algebra {\Large \pounds }.
Since the commutation relations are closed among these four elements, the
solution $U(t)$ is assumed to have a factorized form
$$
U(t)\equiv e^{\alpha \left( t\right) \mathbf{p}^2}e^{\beta \left( t\right)
\mathbf{p}}e^{\gamma \left( t\right) \mathbf{x}}e^{\mu (t)}\eqno(a3)
$$
Its Schroedinger equation defines an solvable system of equations about the
time-dependent parameters $\alpha \left( t\right) ,\beta \left( t\right)
,\gamma \left( t\right) $ and $\mu (t):$%
\[
\frac d{dt}\alpha \left( t\right) =-\frac i{2M}
\]
$$
\frac d{dt}\beta \left( t\right) -2i\alpha \left( t\right) \frac d{dt}\gamma
\left( t\right) =\frac{iA}M\eqno(a4)
$$
\[
\frac d{dt}\gamma \left( t\right) =-if
\]
\[
\frac d{dt}\mu \left( t\right) -i\beta \left( t\right) \frac d{dt}\gamma
\left( t\right) =0
\]
The solution is
\[
\alpha \left( t\right) =-\frac{it}{2M}=i\stackrel{\symbol{126}}{\alpha }(t)
\]
$$
\beta \left( t\right) =-\frac{ift^2}{2M}+\frac{iAt}M=i\stackrel{\symbol{126}%
}{\beta }\left( t\right) \eqno(a6)
$$
\[
\gamma \left( t\right) =-itf
\]
\[
\mu \left( t\right) =-\frac{if^2t^3}{6M}+\frac{iAft^2}{2M}=i\stackrel{%
\symbol{126}}{\mu }\left( t\right)
\]

The action of the evolution operator
$$
U_k(t)\equiv e^{\stackrel{\symbol{126}}{i\alpha }\left( t\right) \mathbf{p}%
^2}e^{\stackrel{\symbol{126}}{i\beta }\left( t\right) \mathbf{p}%
}e^{-iftx}e^{i\stackrel{\symbol{126}}{\mu }(t)}\eqno(a3)
$$
transforms the initial state in momentum representation
$$
\varphi (p,0)=\langle p|\varphi (0)\rangle =(\frac{2a^2}\pi )^{\frac
14}e^{-a^2p^2}\eqno(a3)
$$
into

\
$$
\varphi (p,t)=\langle p|U(t)|\varphi (0)\rangle \newline
=(\frac{2a^2}\pi )^{\frac 14}e^{i\stackrel{\symbol{126}}{\mu }(t)}e^{%
\stackrel{\symbol{126}}{i\alpha }\left( t\right) p^2}e^{\stackrel{\symbol{126%
}}{i\beta }\left( t\right) p}e^{-a^2(p+ft)^2}\eqno(a3)
$$
Then in momentum representation we can easily calculate the overlap of two
entangled states:
\begin{eqnarray*}
|\langle \varphi ^{\prime }(t)|\varphi (t)\rangle | &=&\exp
[-a^2t^2(f^2+f^{\prime 2})]\times \\
&&|\exp \{\frac{-[\stackrel{\symbol{126}}{\beta }\left( t\right) -\stackrel{%
\symbol{126}}{\beta }^{\prime }\left( t\right) +i2a^2t(f+f^{\prime })]^2}{%
8a^2}\}|
\end{eqnarray*}

\newpage
\includegraphics{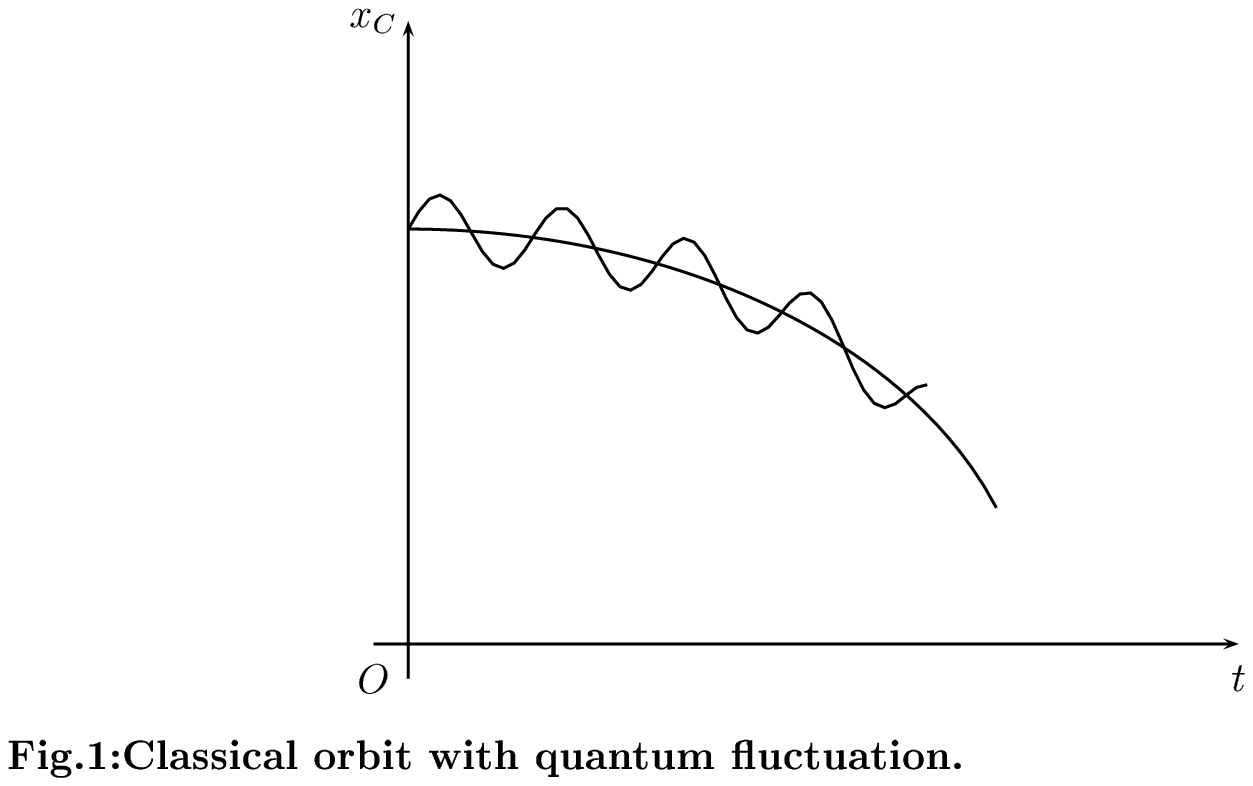}
\newpage
\includegraphics{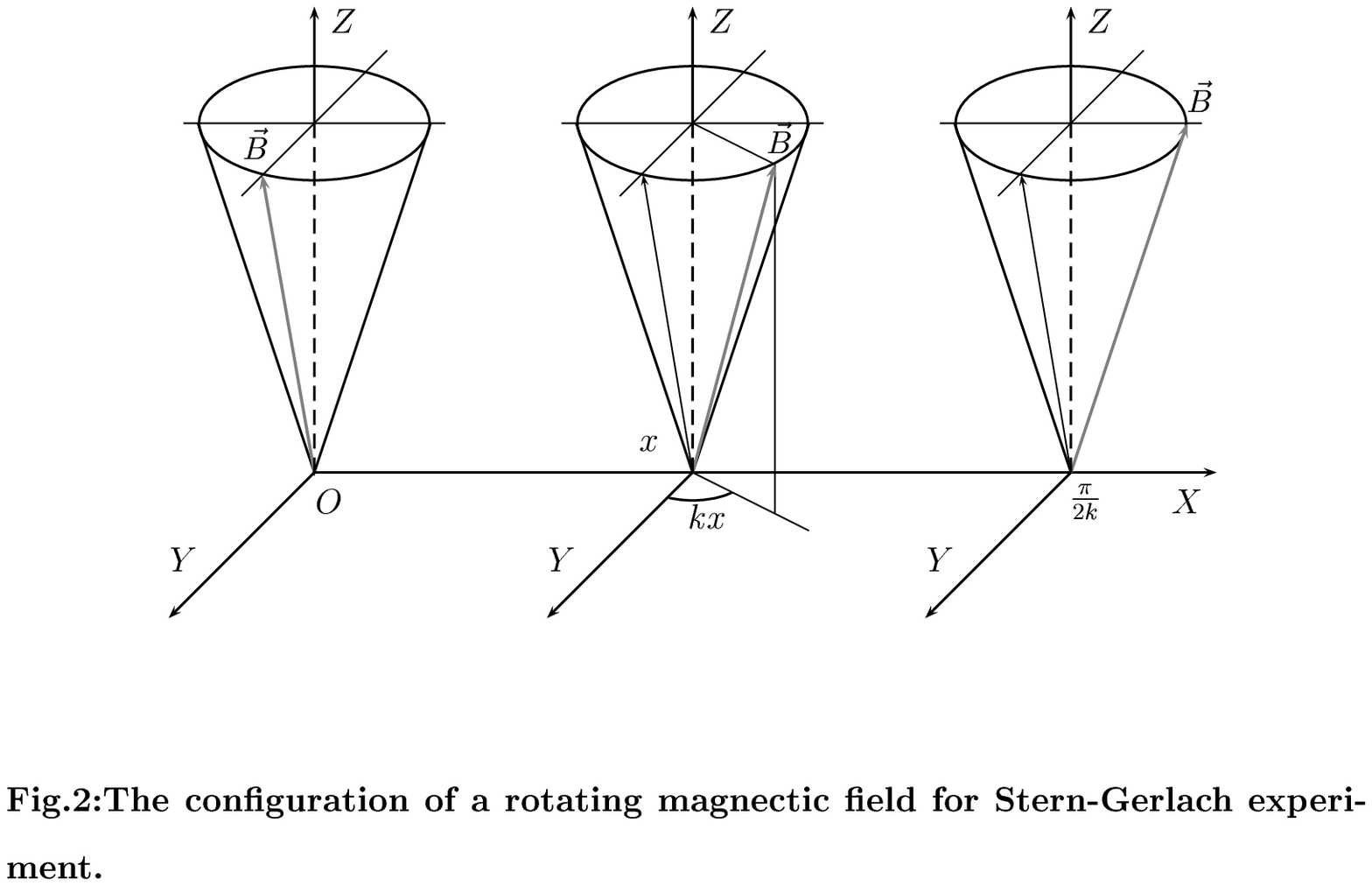}
\newpage
\includegraphics{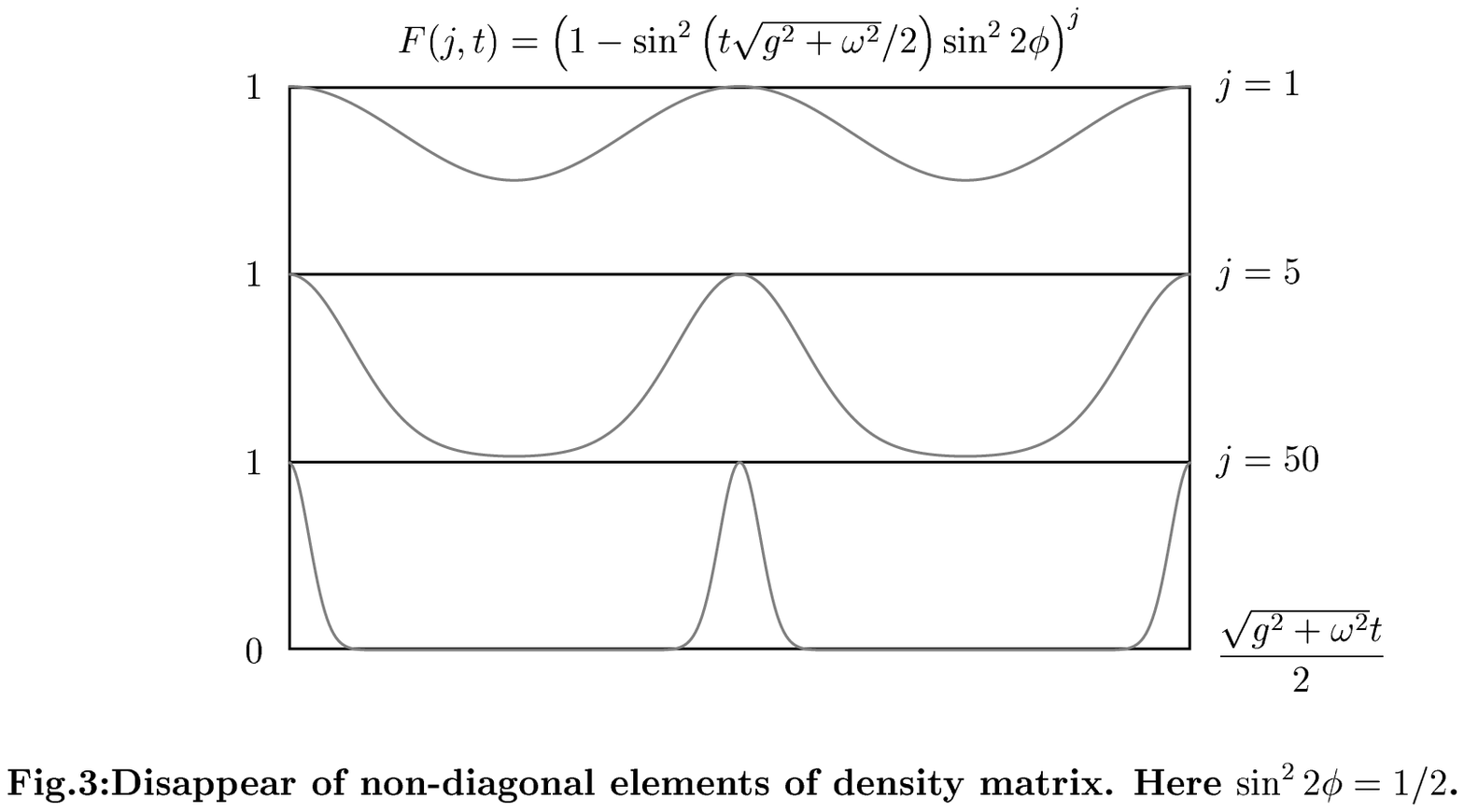}
\newpage
\includegraphics{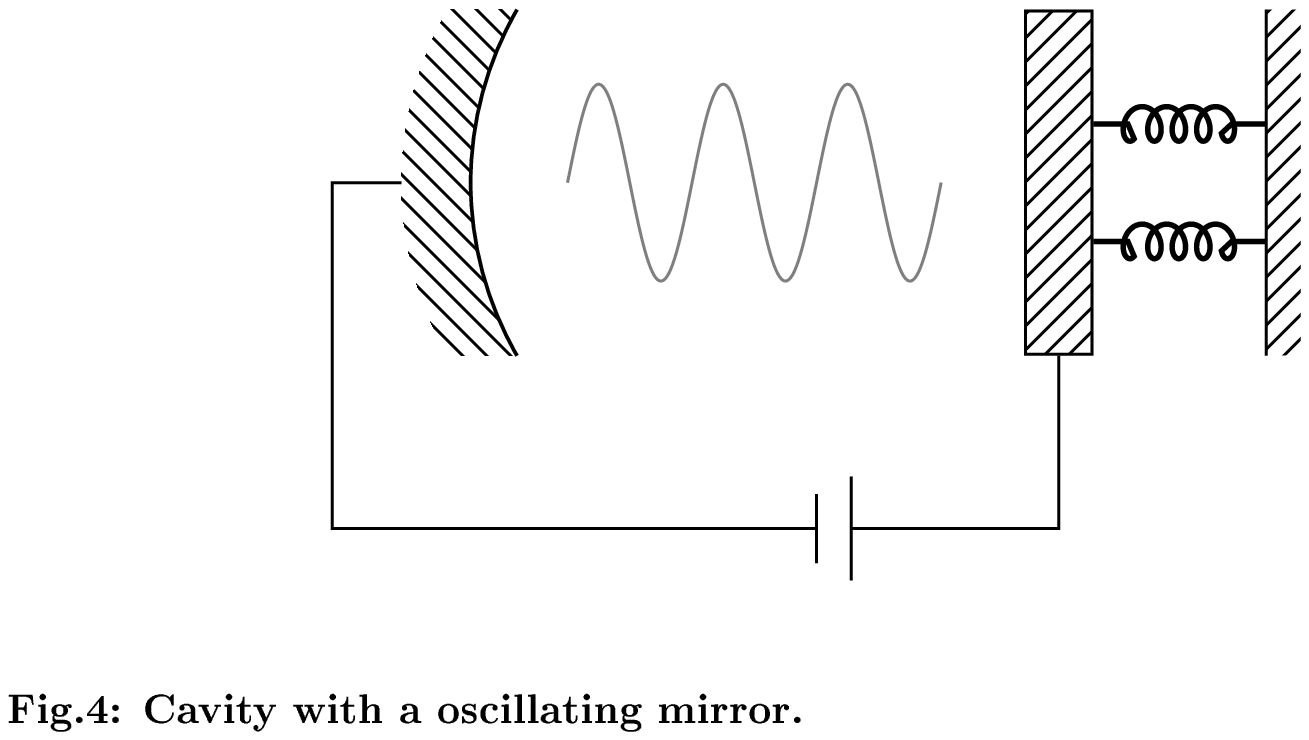}
\newpage
\includegraphics{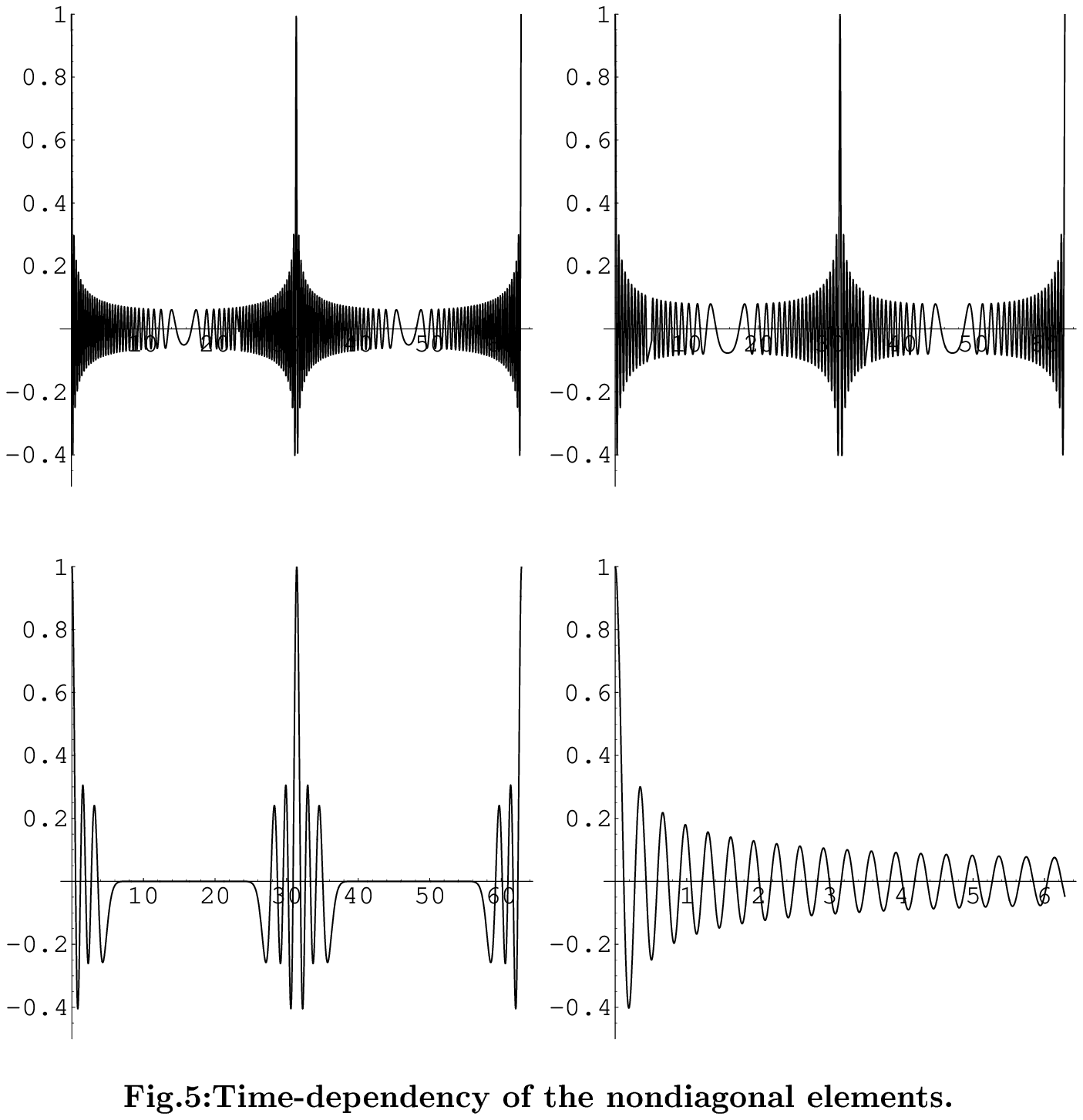}
\newpage
\includegraphics{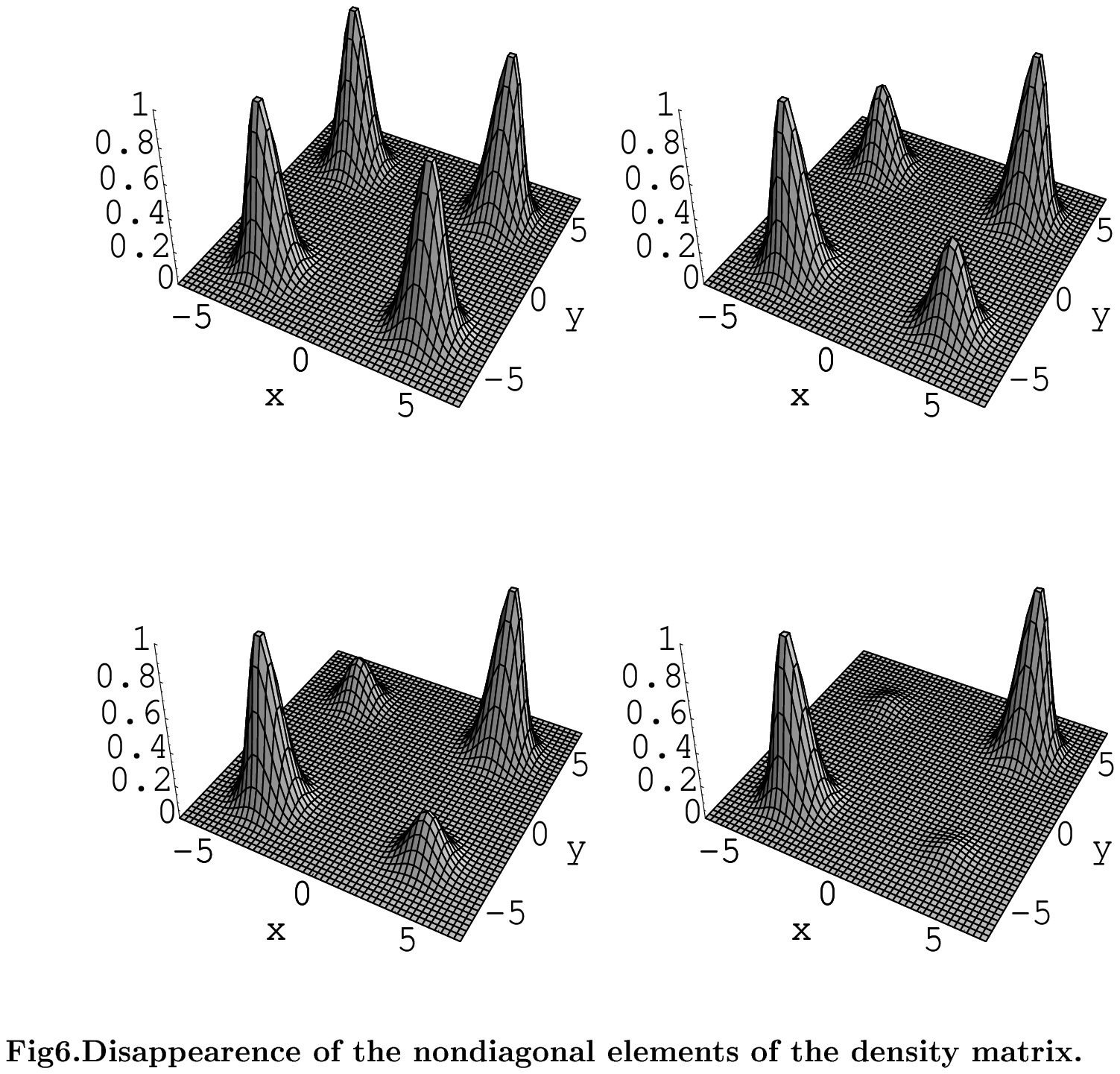}
\newpage
\end{document}